\documentclass[aps,prl,twocolumn,showpacs,floatfix,preprintnumbers,amsfont,amsmath,amssymb,nofootinbib,superscriptaddress]{revtex4-1}

\usepackage{color}
\usepackage{bbold}
\usepackage[colorlinks=true,linktocpage=true,linkcolor=colorLink,citecolor=colorCite,urlcolor=colorURL]{hyperref}
\usepackage[normalem]{ulem}
\usepackage{lipsum}
\usepackage{url}
\usepackage{float}
\usepackage[table]{xcolor}
\usepackage{graphicx}
\usepackage[font=small, justification=raggedright, format=plain]{caption}
\usepackage{subcaption}
\usepackage{mathtools}
\usepackage{bm}
\usepackage{siunitx}
\usepackage{multirow}
\definecolor{colorLink}{rgb}{0.9,0,0} % red
\definecolor{colorCite}{rgb}{0,0.7,0} % green
\definecolor{colorURL} {rgb}{0,0,0.8} % navy

\usepackage{array,etoolbox}
\preto\tabular{\setcounter{magicrownumbers}{-1}}
\newcounter{magicrownumbers}

\colorlet{RED}{red}

\def\chieff{\chi_{\rm eff}}

\def\msun{{\rm M_{\odot}}}
\newcommand{\sk}[1]{}
\usepackage{fontawesome}

\newcommand{\be}{\begin{equation}}
\newcommand{\ee}{\end{equation}}
\newcommand{\ba}{\begin{eqnarray}}
\newcommand{\ea}{\end{eqnarray}}

\newcommand{\apjl}{Astrophys. J. Lett.}

\newcommand{\mnras}{Mon. Not. R. Astron. Soc.}

\allowdisplaybreaks
\begin{document}

\title{Significant increase in sensitive volume of a gravitational wave search\\ upon including higher harmonics}

\author{Ajit Kumar Mehta}
% \email{ajit\_mehta@ucsb.edu}
\affiliation{\mbox{Department of Physics, University of California at Santa Barbara, Santa Barbara, CA 93106, USA}}
\author{Digvijay Wadekar}
\affiliation{\mbox{Department of Physics and Astronomy, Johns Hopkins University,
3400 N. Charles Street, Baltimore, Maryland, 21218, USA}}
\affiliation{\mbox{School of Natural Sciences, Institute for Advanced Study, 1 Einstein Drive, Princeton, NJ 08540, USA}}
\author{Javier Roulet}
\affiliation{Theoretical AstroPhysics Including Relativity and Cosmology, California Institute of Technology, Pasadena, California, USA}
\author{Isha Anantpurkar}
\affiliation{\mbox{Department of Physics, University of California at Santa Barbara, Santa Barbara, CA 93106, USA}}
\author{Tejaswi Venumadhav}
\affiliation{\mbox{Department of Physics, University of California at Santa Barbara, Santa Barbara, CA 93106, USA}}
\affiliation{\mbox{International Centre for Theoretical Sciences, Tata Institute of Fundamental Research, Bangalore 560089, India}}
\author{Jonathan Mushkin}
\affiliation{\mbox{Department of Particle Physics \& Astrophysics, Weizmann Institute of Science, Rehovot 76100, Israel}}
\author{Barak Zackay}
\affiliation{\mbox{Department of Particle Physics \& Astrophysics, Weizmann Institute of Science, Rehovot 76100, Israel}}
\author{Matias Zaldarriaga}
\affiliation{\mbox{School of Natural Sciences, Institute for Advanced Study, 1 Einstein Drive, Princeton, NJ 08540, USA}}
\author{Tousif Islam}
\affiliation{Kavli Institute for Theoretical Physics, University of California Santa Barbara, Santa Barbara, CA 93106}
\affiliation{Theoretical AstroPhysics Including Relativity and Cosmology, California Institute of Technology, Pasadena, California, USA}
\date{\today}

%%%%%%%%%%%%%%%%%%%%%%%%%%%%%%%%%%%%%%%%%%%%%%%%%%%%%%%%%%%%%%%%%%%%%%%%%%%%%%%

\begin{abstract}
Nearly all gravitational wave searches to date have included only the quadrupole mode in their search templates. Here, we demonstrate that incorporating higher harmonics improves the search sensitive volume for detecting binary black hole mergers, challenging the conclusion of previous studies. 
Using the $\tt{IAS-HM}$ detection pipeline\footnote{\label{foot}The $\tt{IAS-HM}$ pipeline is publicly available at \url{https://github.com/JayWadekar/gwIAS-HM}} \cite{Wad23_Pipeline}, and the simulated (injection) signals from the LIGO-Virgo-KAGRA (LVK) collaboration, we quantify the improvement in sensitivity due to the inclusion of higher harmonics. This improvement is significant for systems with higher mass ratios and larger total masses, with gains in sensitivity even exceeding $100\%$ at certain high masses. We also show that, due to using a marginalized detection statistic, the $\tt{IAS-HM}$ pipeline performs roughly as well as its quadrupole-mode-only counterpart even for equal mass-ratio mergers, and its sensitive volume is either better than or comparable to that of the individual LVK pipelines. \href{https://zenodo.org/records/14752874}{\faGithub}
%\ajitm{even when the Virgo detector is not included in the IAS pipelines}
\end{abstract}

\maketitle

\section{Introduction}

The LIGO, Virgo and KAGRA detectors \cite{LIGOScientific:2014pky, VIRGO:2014yos, KAGRA:2020tym} currently provide strain data that contain gravitational wave (GW) signals from mergers of compact binary systems, such as binary black holes (BBHs). Multiple GW detection pipelines, developed both within and outside the LVK collaboration \cite{CANNON2021100680,Aubin:2020goo,Klimenko:2015ypf,Usman:2015kfa,ias_pipeline_o1_catalog_new_search_prd2019, ias_o2_pipeline_new_events_prd2020, Ols22_ias_o3a, NitzCatalog_1-OGC_o1_2018, NitzCatalog_2-OGC_o2_2020, nitz_o3a_3ogc_catalog_2021, nitz_4ogc_o3_ab_catalog_2021,Chi23,Meh23_ias_o3b, Kol24_Ares_ML_Search,Wad23_Pipeline}, analyze this data to identify the signals. In the first three observing runs, approximately 100 compact binary mergers have been detected \cite{lvc_gwtc3_o3_ab_catalog_2021, nitz_4ogc_o3_ab_catalog_2021, Wad23_HM_Events, Meh23_ias_o3b}. These detections have significantly advanced our understanding of astrophysics \cite{LVKO3bpopulation, ias_o3a_population_analysis_prd2021roulet, nitz_4ogc_o3_ab_catalog_2021} and fundamental physics  \cite{LIGOScientific:2021sio}.

Most detections to date have been made using search templates that include only the dominant harmonic (the $(\ell, |m|) = (2, 2)$ mode) of GW signals \cite{CANNON2021100680,Aubin:2020goo,Klimenko:2015ypf,Usman:2015kfa,ias_pipeline_o1_catalog_new_search_prd2019, ias_o2_pipeline_new_events_prd2020, Ols22_ias_o3a, NitzCatalog_1-OGC_o1_2018, NitzCatalog_2-OGC_o2_2020, nitz_o3a_3ogc_catalog_2021, nitz_4ogc_o3_ab_catalog_2021,Chi23,Meh23_ias_o3b}.
%, primarily due to computational constraints. 
However, signals with substantial contributions from higher modes (HM), such as those emitted by BBH systems with high mass ratios or high total masses (particularly in the intermediate-mass black holes (IMBH) range), may be missed by these $(2, 2)$-mode-only pipelines \cite{Wad23_TemplateBanks, HMeffect_ParameterSpaceDependency_PekowskyPRD2013, HMeffect_RelativeModeSignificance_HealyPRD2013,  Cap14, HMeffect_AlignedSearchImpactCalderonBustilloPRD2016, HMandPrecessionEffect_HeavySearchImpact_CalderonBustilloPRD2017, HMeffect_IMBHsearchImpact_CalderonBustilloPRD2018, Mil21, Har18, Cha22, Sha22,Zha23,LVK_O3_IMBH_search}. Detecting such systems can significantly improve our knowledge of the diverse formation channels of BBHs \cite{doi:10.1142/S0218271804004426,2011GReGr..43..485G, GW190412, GW190814, hierarchical_7merger_scenario2020b, hierarchical_from_dynamical_in_any_star_cluster2020b, hierarchical_mergerFamily_dynamical_mass_dist_matters2021, hierarchical_rate_sensitive_to_natal_spins_Fragione2021kocsis, hierarchical_mergers_agn_kocsis2019,agn_bbh_population_chieff_q_simulation_mckernan_ford2019, bbh_spin_evolution_agn_Tagawa_2020a, bbh_evolution_agn_merger_timescale_ishibashi2020a, migration_traps_spins_rates_mckernan_ford2020a, 
mass_gap_agn_bbh_mergers2021, agn_accretion_disk_merger_population2020a, Tag20_AGN, Sam22_AGN}, and physics of the pair-instability mass gap \cite{Marchant:2020haw, Farmer:2019jed, VanSon20_UMG_Pollution, Mehta:2021fgz, Hen23_UMG, Gol23_UMG, Fra24_UMG}.
The lack of IMBH detections at the moment leaves their formation mechanisms highly uncertain and poorly understood. \cite{Fra18_IMBH_LISA, Kri23_NSC, Kri24_GasAccretion, Kri24_Rapster}.

Incorporating HM in search templates introduces two competing effects: on one hand, it increases the captured signal-to-noise ratio (SNR), while on the other hand, it causes an increase in background noise and degradation of false-alarm rate (FAR) of detected signals (as noise can increasingly mimic real signals due to addition of extra degrees of freedom introduced by HM in the templates) \cite{Cap14, Cha20_VT, Cha22}. Due to this trade-off, earlier studies concluded that including HMs in searches was unlikely to enhance the overall detection sensitivity of compact binary mergers \cite{Cap14, Cha22}. 

We recently developed a new efficient \emph{mode-by-mode filtering} method to include HM into a search pipeline (see \cite[figure~1]{Wad23_Pipeline}). We matched-filter the strain data with different normalized harmonics and then combine the resulting signal-to-noise timeseries \cite{Wad23_TemplateBanks}. This has two advantages: $(i)$ the matched-filtering cost becomes linearly proportional to the number of HM used (instead of increasing by a factor of $\sim 100$ 
if the waveform templates are used to model all harmonics at once \cite{Har18, Cha22}); $(ii)$ it allows for efficient marginalization over the extra degrees of freedom corresponding to HM (namely the binary inclination and the initial reference phase). It is worth noting that the mode-by-mode filtering method can also be used for efficient parameter estimation of the detected binaries \cite{Fai23_SimplePE, Rou23_CoherentScore}. We leave the details of our detection statistics and template banks to Refs.~\cite{Wad23_Pipeline} and \cite{Wad23_TemplateBanks} respectively.

Our goal in this paper is to precisely quantify increase/decrease in the sensitive volume upon introducing HM (using a mode-by-mode filtering method) in the search pipeline. We estimate the sensitive volume of the $\tt{IAS-HM}$ pipeline$^{\ref{foot}}$ by analyzing simulated signals (injections) made publicly available by the LVK collaboration on Zenodo corresponding to the second part of the third observing run (O3b) \cite{zenodoLVK}. We also estimate the sensitivity of the $\tt{IAS-HM}$ pipeline in its $(2, 2)$-mode-only limit (referred to as the $\tt{IAS-22}$ pipeline). This allows us to understand the improvement in sensitivity volume solely due to the inclusion of HM.

We had performed the first search accounting for HM throughout the BBH parameter space in Ref.~\cite{Wad23_HM_Events} (see also Ref.~\cite{Cha22}). 
Using the $\tt{IAS-HM}$ pipeline, we had found $\sim$ 10 new BBH mergers, most of which had median primary masses\footnote{The heavier masses of the binary components.} exceeding $60\, M_{\odot}$ in the source frame \cite{Wad23_HM_Events}. As we discuss below, this mass range aligns with where we observe significant improvements in detection sensitivity volume upon including HM, adding more credibility to our previously reported detections.

% We will demonstrate that incorporating HM leads to a significant improvement in detection sensitivity across the BBH and IMBH parameter space.
% The $\tt{IAS-HM}$ pipeline includes HMs with a computational cost proportional to the number of HMs used, making it much more efficient than previous HM-based searches. The search identified $\sim 10$ new BBH merger events, most of which had median primary masses exceeding $60\, M_{\odot}$ in the source frame. 

\begin{figure*}[t]
    \centering
    \includegraphics[width=0.99\textwidth]{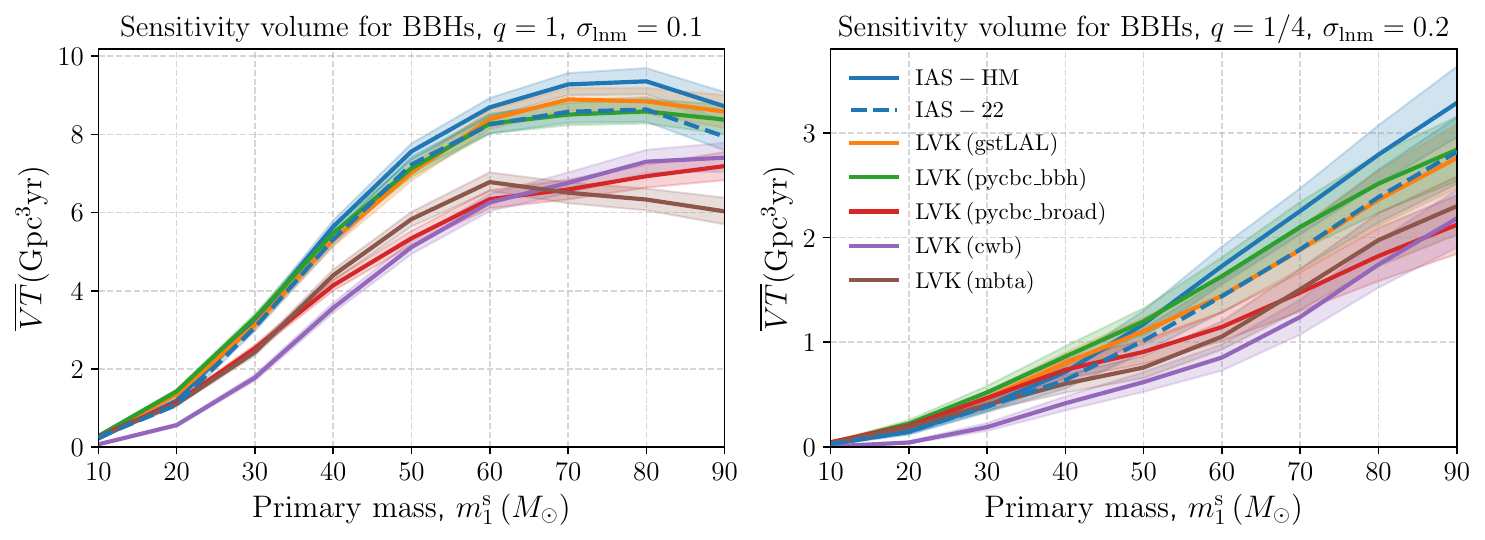}
    \vspace{-0.2cm}
    \caption{Detection sensitivity volume and its measurement uncertainty for the $\tt{IAS}$ and LVK search pipelines, using a detection criterion of $p_{\rm{astro}}>0.5$.
    The $x$-axis represents the mean values of the log-normal distributions assumed for the primary masses in the source frame (see Eq.~\eqref{eq:lognormal}).
    \textbf{Left panel}: For $q=1$. The $\tt{IAS}$ pipeline with HM ($\tt{IAS-HM}$) performs equally well as its $(2, 2)$-only counterpart ($\tt{IAS-22}$) and demonstrates performance that is roughly comparable to, the individual LVK pipelines. \textbf{Right panel}: For $q=1/4$. The $\tt{IAS-HM}$ pipeline performs significantly better than both the $(2, 2)$-mode counterpart $\tt{IAS-22}$, as well as the individual LVK pipelines towards the higher masses. This is expected for lower mass ratios and larger total masses, where the contributions from HM to the signals become increasingly significant. 
    %The LVK-$\tt{Any}$ case is shown separately in dotted as it is just an optimistic estimate which combines only the detected injections from all the LVK pipelines (without accounting for an increase in the overall background due to the look-elsewhere effect), see the text for details. 
    Note that, for both panels, the IAS pipelines were run only on Hanford and Livingston detectors (including Virgo will further improve the $\tt{IAS}$ $\overline{VT}$ estimates).}
    \label{fig:vol_q_1} 
\end{figure*}

Our results are in Figs.~\ref{fig:vol_q_1} and~\ref{fig:vol_q_1_imbh}. For low-mass BBH systems, the sensitivity either remains roughly the same or shows a modest improvement. The improvement becomes more pronounced as the mass ratio and primary mass (or the total mass) increase. For reference, for IMBH binaries with a mass ratio $q=1/4$, we find that HM enhance the sensitivity by $\sim 15\%-150\%$ for primary mass between $\sim [100, 250]\, M_\odot$.
%Addressing the open question of the IMBH mass function through GW observations could provide transformative insights into various astrophysical phenomena, such as unraveling the processes involved in the formation of supermassive black holes \cite{Ebisuzaki_2001}.

% The remainder of this paper is organized as follows: in Section \ref{sec:method}, we describe the method used to estimate the sensitivity volume and its measurement uncertainties. Section \ref{sec:result} presents our results for BBH and IMBH binaries. It also discusses the gains achieved in the sensitivity volume due to HM. Section \ref{sec:conclusion} summarizes our conclusions and outlines future directions.

\section{Method}
\label{sec:method}
The expected number of detections, $N_{\rm{det}}$, from a given pipeline can be expressed as:
\begin{equation}
    N_{\rm{det}} = \displaystyle \int \Bigg(\dfrac{d^3 N}{dt_s dV_c d\theta}\Bigg )\, \dfrac{dV_c}{dz}\, \dfrac{dt_s}{dt}\, p(\mathrm{det}|\theta, z)\,  dt d\theta dz
\label{eq:Ndet}
\end{equation}
Here, the quantity inside the parentheses describes the number of compact binary mergers per unit source-frame time ($dt_s$), per unit comoving volume ($dV_c$) for a given set of intrinsic source parameters $\theta = (m_1^{\mathrm{s}}, m_2^{\mathrm{s}}, \Vec{s}_1, \Vec{s}_2)$. The parameters $m_{1,2}^{^{\mathrm{s}}}$ denote the primary and secondary mass of the compact binary merger in the source-frame with $m_2^{\mathrm{s}} \leq m_1^{\mathrm{s}}$, and $\Vec{s}_{1,2}$ represent the component spin vectors. The term $p(\mathrm{det}|\theta, z)$ denotes the probability that a merger signal with parameters $(\theta, z)$ is detected by the considered pipeline. The observer-frame time ($dt$) relates to the source-frame time ($dt_s$) via $dt = (1+z)\,dt_s$, accounting for cosmic time dilation.

The term in parentheses can be further decomposed as:
\begin{equation}
    \dfrac{d^3 N}{dt_s dV_c d\theta} = R(z)\, p(\theta)
\end{equation}
where $R(z)$ is the merger rate density as a function of redshift, and  $p(\theta)$ is the astrophysical probability distribution over the intrinsic source parameters $\theta$.  The merger rate density can be parameterized as $R(z) = R_0 \, f(z)$,
where $R_0$ is the local merger rate density, i.e, at $z=0$ and $f(z)$ accounts for the redshift evolution of the rate density. 

Substituting these definitions into Eq.~\eqref{eq:Ndet}, the expected number of detections can be written as:
\begin{equation}
     N_{\rm{det}} = R_0\, \int f(z)\, \dfrac{1}{1+z}\,\dfrac{dV_c}{dz}\, p(\theta)\, p(\mathrm{det}|\theta, z)\,  dt d\theta dz
\end{equation}

The {detection sensitivity volume} is defined as,
\begin{equation}
    \overline{VT} = \int f(z)\, \dfrac{1}{1+z}\,\dfrac{dV_c}{dz}\, p(\theta)\, p(\mathrm{det}|\theta, z)\,  dt d\theta dz
\label{eq:VT}
\end{equation}
so that $N_{\rm{det}} = R_0\,  \overline{VT}$.
Thus,  $\overline{VT}$ represents the comoving volume-time accessible to the detector, weighted by astrophysical, cosmological, and detector-specific factors.

To compute  $\overline{VT}$ from Eq.~\eqref{eq:VT}, we use a Monte Carlo integral over found injections \cite{Tiwari:2017ndi}:

\begin{equation}
     \overline{VT} \simeq \dfrac{T_{\mathrm{obs}}}{N_{\mathrm{draw}}} \sum_{j=1}^{N_{\mathrm{found}}} \dfrac{f(z_j)\, \dfrac{1}{1+z_j}\,\dfrac{dV_c(z_j)}{dz}\, p(\theta_j)}{{\pi_{\mathrm{draw}}(\theta_j, z_j)}}
\label{eq:VT_est}
\end{equation}
where $T_{\mathrm{obs}}$ is the observation time of the dataset analyzed,  $N_{\mathrm{draw}}$ is the number of injections generated from the sampling distribution 
$\pi_{\mathrm{draw}}$, and $N_{\mathrm{found}}$  is the number of injections detected by the pipeline. The Monte Carlo estimate weights each found injection by the inverse of the sampling probability $\pi_{\mathrm{draw}}$, ensuring unbiased sensitivity volume estimates. 

The uncertainty in $\overline{VT}$ is evaluated using the bootstrap method. Specifically, {we resample the injections with replacement, generating multiple bootstrap samples of size $N_{\mathrm{draw}}$}. For each bootstrap sample, we compute the sensitivity volume. The mean of these bootstrap samples is used as the final estimate of $\overline{VT}$, while the $90\%$ confidence interval from the bootstrap distribution provides the error bars. This approach ensures a robust characterization of the uncertainty in the sensitivity volume. In all results presented in this paper, we use the bootstrap mean as the representative value of $\overline{VT}$, and the $90\%$ confidence interval as the associated error bar. As we would expect, $\overline{VT}$ closely follows a normal distribution with a variance that can be expressed as:
\begin{equation}
    \sigma^2 = \dfrac{\overline{VT}^2}{N_{\mathrm{eff}}}
    \label{eq:sigma_VT}
\end{equation}
where $N_{\mathrm{eff}}$ is the effective number of detected injections, accounting for the weighting by $\pi_{\mathrm{draw}}$. An expression for $N_{\mathrm{eff}}$ can be found in Ref.~\cite[equation~9]{Farr_2019}.

% The $x$-axis, in this case, also denotes half the mean total mass of the BBH mergers.
% \begin{figure}[t]
%     \centering
%     \includegraphics[width=0.48\textwidth]{plots/sens_vol_q0.25.pdf}
%     \vspace{-0.2cm}
%     \caption{Similar to  Fig.~\ref{fig:vol_q_1} with $q=1/4$. A larger width ($\sigma_{\mathrm{lnm}}=0.2$) is used to ensure a significant $N_{\mathrm{eff}}$ and thus, the error in the sensitivity volume estimates. The $\tt{IAS-HM}$ pipeline, which incorporates HM, performs better than both the $(2, 2)$-mode counterpart $\tt{IAS-22}$ as well as the individual LVK pipelines towards the higher masses when compared to the case of $q=1$ in Fig.~\ref{fig:vol_q_1}. This is expected for higher mass ratios and larger total masses, where the contributions from HM to the signals become increasingly significant.}
%     \label{fig:vol_q_4}
% \end{figure}

In this work, we calculate $\overline{VT}$ for the $\tt{IAS}$ pipelines and compare it with the $\overline{VT}$ of various LVK pipelines. We use the injection datasets made publicly available by the LVK collaboration on Zenodo \cite{zenodoLVK}. We focus on the observing run O3b throughout this paper. The injection datasets include:
\begin{itemize}
    \item A BBH injection set, with source-frame component masses in the range $[2, 100]\, M_{\odot}$. The primary mass follows a power-law distribution, $p(m_1^s) \propto (m_1^s)^\alpha$, where the exponent $\alpha = -2.35$. The secondary mass ($m_2^s$) is drawn from a conditional power-law distribution with exponent $\alpha = 1$, conditioned on $m_1^s$. Spins are isotropically distributed in direction and uniformly in magnitude. The redshift distribution in Eq.~\eqref{eq:VT} is given by
    $f(z) = 1+z$ with a maximum redshift $z_{\mathrm{max}} = 1.9$.
    
    \item  An IMBH binary injection set, with primary masses between $[90, 600]\, M_{\odot}$ and secondary masses in the range $[10, 600]\, M_{\odot}$. Both the primary mass distribution, $p(m_1^s)$, and the conditional distribution of the secondary mass, $p(m_2^s | m_1^s)$, follow a power-law with exponent $\alpha = -1$. The spin distribution is the same as the BBH injection set. The redshift distribution assumes no evolution with $f(z)=1$, and a maximum redshift $z_{\mathrm{max}} = 2.5$.

\end{itemize}

For each injection dataset, we evaluate $\overline{VT}$ of the $\tt{IAS}$ pipelines by injecting signals into the O3b data from the Hanford and Livingston detectors. {We use $\tt{IMRPhenomXPHM}$ approximant \cite{Pratten:2020ceb} to compute the polarizations of the injected signals}. For the BBH injection set, we apply the detection criterion $p_{\mathrm{astro}}>0.5$ (where $p_\mathrm{astro}$ refers to the probability of a trigger being astrophysical in origin, as opposed to a noise artifact), consistent with thresholds used in real LVK and $\tt{IAS}$ searches. For the IMBH dataset, however, only IFAR (i.e., $\mathrm{FAR}^{-1}$) values are provided (and not $p_{\mathrm{astro}}$), and furthermore, only the results from one of the LVK pipelines ($\tt{pycbc\_bbh}$) are reported. Therefore, for the IMBH dataset, we use $\mathrm{IFAR}>1$ year  as the detection criterion and compare our results only with $\tt{pycbc\_bbh}$.

 It is important to note that the $\tt{IAS}$ pipelines currently analyze data from only from the Hanford and Livingston detectors, whereas LVK pipelines incorporate Virgo data as well. Including additional detectors enhances sensitivity volume. Thus, while comparing $\overline{VT}$ between the $\tt{IAS}$ and LVK pipelines, this is an important point to keep in mind. We aim to include also the Virgo data in future IAS analysis papers.

A key objective of this work is to evaluate the improvement in $\overline{VT}$ due to HM. To achieve this, we compute the ratio ${\overline{VT}_{\mathrm{HM}}}/{\overline{VT}_{22}}$, also expressed as the fractional gain:
\begin{equation} 
\Delta \overline{VT} = \dfrac{\overline{VT}_{\mathrm{HM}}}{\overline{VT}_{22}} - 1 
\end{equation}
where $\overline{VT}_{\mathrm{HM}}$ is the sensitivity volume of the $\tt{IAS-HM}$  pipeline, and  $\overline{VT}_{22}$ corresponds to the $\tt{IAS-22}$ pipeline. We estimate the uncertainty in $\Delta \overline{VT}$ using the bootstrap method as explained before. By systematically evaluating $\Delta \overline{VT}$ across different masses (as discussed in the following section), we identify regions of the parameter space where the inclusion of higher harmonics significantly enhances the sensitivity volume.

\section{Results and Discussion}
\label{sec:result}
\subsection{Sensitivity volume for BBH injection set}
% In dotted gray, we show the results from naively picking coincident candidates above $p_\mathrm{astro}>0.5$ in any of the LVK pipelines. Note however that for a 
% consistent statistical analysis, background from each search should also be included, see e.g., \cite{Ban23_Unified_pastro}
In Fig.~\ref{fig:vol_q_1}, we present the sensitivity volume ($\overline{VT}$) and its associated measurement uncertainties for the $\tt{IAS}$ and LVK pipelines as a function of the primary source-frame mass ($m_1^{\mathrm{s}}$). For this analysis and throughout the work, we assume a log-normal distribution for the primary and secondary component masses in the astrophysical prior $p(m_1^{\mathrm{s}}, m_2^{\mathrm{s}})$ in Eq.~\eqref{eq:VT_est}, consistent with the LVK  analysis in Ref.~\cite{lvc_gwtc3_o3_ab_catalog_2021}.
Specifically,

\begin{widetext}
\begin{equation}
p(m_1^{\mathrm{s}}, m_2^{\mathrm{s}}) =
\begin{cases} 
\dfrac{c}{2\pi \sigma_{\text{lnm}}^2  m_1^{\mathrm{s}} m_2^{\mathrm{s}}} 
\exp\Bigg(-\dfrac{1}{2} \dfrac{(\log m_1^{\mathrm{s}} - \log M_1)^2}{\sigma_{\text{lnm}}^2}\Bigg) 
\exp\Bigg(-\dfrac{1}{2} \dfrac{(\log m_2^{\mathrm{s}} - \log M_2)^2}{\sigma_{\text{lnm}}^2}\Bigg), & \text{for } m_{2}^{\mathrm{s}} < m_{1}^{\mathrm{s}}, \\
0, & \text{otherwise.}
\end{cases}
\label{eq:lognormal}
\end{equation}
\end{widetext}
For spins and other parameters, we adopt the same distributions as the injection set, i.e., $\pi_{\text{draw}}$.

The $x$-axes in Fig.~\ref{fig:vol_q_1} represent the mean values ($M_1$) of the log-normal distribution for the primary mass. We use a fixed variance of $\sigma_\mathrm{lnm}=0.1$ in the left panel. In the right panel, however, we use a larger width ($\sigma_{\mathrm{lnm}}=0.2$) to maintain a sufficiently large $N_{\mathrm{eff}}$, ensuring the smoothness of curves as well as upper bounds on the error bars. In the left panel, the mean mass ratio ($q =M_2/M_1$) is held constant at 1, so the $x$-axis also corresponds to half the mean total source-frame mass. This figure is derived from the BBH injection set, which spans component masses up to $100\, M_{\odot}$. To account for this hard cut-off, we appropriately re-normalize the log-normal mass distributions. To ensure sufficient statistical precision of $\overline{VT}$, characterized by $N_{\mathrm{eff}}$, we restrict the $x$-axis to below $90\, M_{\odot}$.
We use $p_\mathrm{astro}>0.5$ as the detection criterion following the approach of other pipelines \cite{lvc_gwtc3_o3_ab_catalog_2021}.
%The secondary masses are similarly modeled with log-normal distributions, where their mean values are given by mass ratio ($q$) $\times$ the primary masses.

\begin{figure*}[t]
    \centering
    \includegraphics[width=0.99\textwidth]{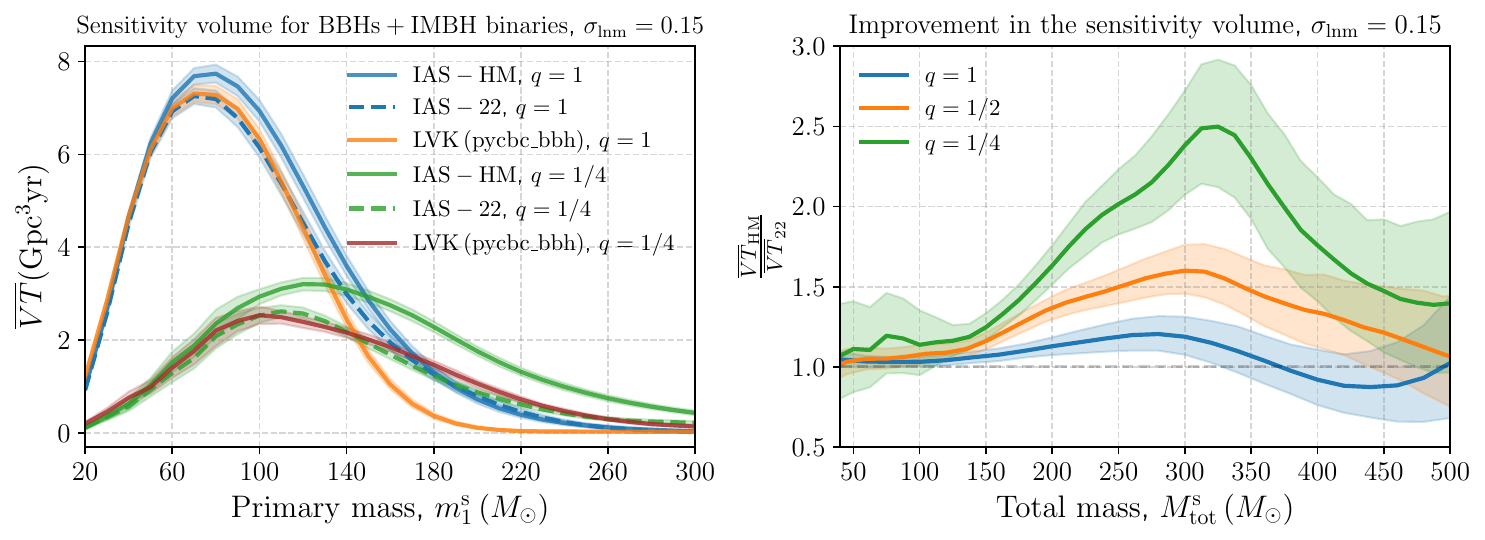}
    \vspace{-0.2cm}
    \caption{Same as Fig.~\ref{fig:vol_q_1}, but including the injection set for IMBH binaries in addition to the lower-mass BBH set. \textbf{Left panel}: Detection sensitivity volumes for two cases: $q=1$ and $q=1/4$. The public LVK IMBH binary injection set only contains results for the $\tt{pycbc\_bbh}$ pipeline and does not include $p_\mathrm{astro}$ values.
    We thus use IFAR$>1$ as the detectability criterion for both the injection sets in this plot and only compare with $\tt{pycbc\_bbh}$. The $\tt{IAS-HM}$ pipeline has comparable or higher sensitivity compared to $\tt{IAS-22}$ throughout the parameter space. Interestingly, for $m_1^s \gtrsim 150 M_\odot$, $q<1$ systems have a comparatively larger observable volume than $q=1$. This makes it relatively more important to include HM for detecting these high-mass IMBH systems. \textbf{Right panel}:  Improvement in the sensitivity volume with the $\tt{IAS-HM}$ pipeline compared to $\tt{IAS-22}$ pipeline for different $q$ against the source-frame total mass. The dashed horizontal line at ${\overline{VT}_{\mathrm{HM}}}/{\overline{VT}_{22}}=1$ indicates no gain or loss in the sensitivity volume. Note that the improvement exceeds $100\%$ (i.e., by a factor of 2) around $M_\mathrm{tot}^s \sim 320\, M_{\odot} $ ($m_1^s\sim 250\, M_{\odot}$) for $q=1/4$.}
    \label{fig:vol_q_1_imbh}
\end{figure*}

\emph{Overall trends:}
Let us first discuss the overall trends seen across all the pipelines.
All the pipelines exhibit an initial increase in $\overline{VT}$ with the total mass of the binary mergers, followed by a plateau, and an eventual decline at higher masses (which is seen more clearly in Fig.~\ref{fig:vol_q_1_imbh}). This behavior is primarily due to the fact that the amplitude of GW signals increases with the total mass, enhancing detectability. For reference, the amplitude scales as $|h|\propto \mathcal{M}^{5/6}$ in the inspiral-dominated regime, where $\mathcal{M} = M_\mathrm{tot} (\frac{q}{(1+q)^2})^{3/5}$ is the chirp mass. However, beyond a certain total mass ($\sim 80\, M_{\odot}$ for most pipelines), the dominant GW frequencies shift out of the detector's sensitive band, reducing the effective detection volume. Interestingly, this flattening is not observed in the right panel of Fig.~\ref{fig:vol_q_1}, which presents results for $q=1/4$. At lower mass ratios, the increased signal duration (and the corresponding higher number of cycles) allows the GW signals to remain within the detector's sensitive frequency band for larger primary masses  (for reference, roughly, $f_\mathrm{merger}\propto M_\mathrm{tot}^{-1}$ for a non-spinning case and thus goes up with lower $q$). Nevertheless, the $\overline{VT}$ for $q=1/4$ is generally lower than that for $q=1$, as the GW signal strength decreases with lower mass ratios. These observations underscore the complex interplay between signal morphology, mass ratio, and detector sensitivity in determining detection volumes across varying parameter spaces.

\emph{Comparison of $(2,2)$+HM and $(2, 2)$-only pipelines:}
The $\tt{IAS-HM}$ pipeline includes the modes: $(2,2)$, $(3,3)$ and $(4,4)$.
The $(2,2)$-only limit of the pipeline is labeled as $\tt{IAS-22}$. 
%This allows us to understand the improvement in sensitivity volume solely due to the inclusion of HM. 
It is worth noting that the $\tt{IAS-22}$ case is an improved version of the $(2, 2)$-only $\tt{IAS}$ pipeline used in our previous searches in \cite{Meh23_ias_o3b, Ols22_ias_o3a}, see the supplemental material for more details.

For $q=1$, $\tt{IAS-HM}$ case exhibits roughly similar sensitivity as $\tt{IAS-22}$. This result is noteworthy because earlier studies have suggested that the inclusion of HM might lead to diminished sensitivity for $q\sim 1$ due to an increased FAR \cite{Cap14}. This worry primarily stemmed from the fact that $q\sim 1$ systems do not gain much in SNR due to HM, but the searches catch additional background due to the additional degrees of freedom in the HM templates. HM do boost the detection of nearly edge-on and asymmetric mass-ratio systems, but such systems have a much smaller observable volume to begin with, and thus the $\overline{VT}$ of the overall search could be compromised.

We designed our marginalized detection statistic in Refs. \cite{Wad23_Pipeline, Rou23_CoherentScore} to limit the increase in search background due to HM.
For example, while performing the marginalization integral over inclination, we add smaller weights to the edge-on configurations instead of face-on configuration (to account for the difference in their observable volume). Similarly, the different binary mass-ratio configurations are also weighed by their observable volume. This helps in preserving the sensitivity of the search in $q=1$ case. 
% Our findings demonstrate that, at the very least, incorporating HM does not degrade the performance, reinforcing the robustness of the $\tt{IAS-HM}$ pipeline.

For $q=1/4$, the $\tt{IAS-HM}$ pipeline exhibits a marked improvement in sensitivity volume with increasing primary mass, outperforming both the $\tt{IAS-22}$ and other LVK pipelines. This enhancement can be attributed to the growing contribution of HM to the signal's power within the detector's sensitive frequency band as the primary mass increases \cite[figure~3]{Wad23_TemplateBanks}. Beyond $\sim 60\, M_{\odot}$, the median improvement in sensitivity volume reaches approximately $\Delta \overline{VT} \sim 15\%$.
%, and this gain continues to increase for higher masses, as discussed in the following section.

\emph{Comparison of IAS and LVK pipelines:} The $\tt{IAS}$ pipelines for $q=1$ perform comparably to the two main LVK pipelines, $\tt{gstLAL}$ and $\tt{pycbc\_{bbh}}$ (despite the absence of Virgo data in the $\tt{IAS}$ analysis).
It is worth mentioning that the injection set also contains the optimistic LVK-$\tt{Any}$ case, based on injections which clear $p_\mathrm{astro}>0.5$ in either of the LVK pipelines. We leave further discussion of this case to the supplemental material.

% \ajitm{In Fig.~\ref{fig:vol_any} (provided in the appendix), we present a consolidated view of the information in Fig.~\ref{fig:vol_q_1}, including an additional curve labeled as LVK-$\tt{Any}$. This curve represents an optimistic estimate of $\overline{VT}$, calculated using the set of injections with $p_\mathrm{astro}>0.5$ in either of the LVK pipelines.}

% the background triggers from each search should also be factored into the analysis. Ignoring these background contributions artificially improves the detection sensitivity estimates. For example, the unified statistical framework proposed in \cite{Banagiri:2023ztt} provides a way to address this challenge by appropriately accounting for the background across pipelines (see also \cite{Bis12_Combining_Pipelines}). Future work could adopt such methods to enable a more comprehensive and fair comparison of the relative performance of different pipelines.
% \ajitm{expand a bit more on this discussion}.

\subsection{Sensitivity volume for BBH $+$ IMBH injection sets}

We will now discuss the results shown in Fig.~\ref{fig:vol_q_1_imbh} for the injection set containing a combination of BBH and IMBH sets using the ``mixture model" available on Zenodo \cite{zenodoLVK}.
% We show the sensitivity volume for binaries including BHs in the the IMBH mass range.
We considered mean mass ratios $q=1$ and $q=1/4$, and took the width of the log-normal distribution to be $0.15$ to have a healthy $N_{\mathrm{eff}}$ in Eq.~\eqref{eq:sigma_VT}. For the IMBH injection set, only the $\tt{pycbc\_{bbh}}$ results are available among the LVK pipelines, and the analysis provides only the IFAR for its recorded injections. We, therefore, adopt a cut on $\mathrm{IFAR}$ to set the detection criterion for $\overline{VT}$ calculations, enabling a fair comparison with the $\tt{IAS}$ pipelines. We choose $\mathrm{IFAR}>1$ year consistent with the threshold used for the population analysis of BHs in GWTC-3 \cite{LVKO3bpopulation}. 
%The methodological details for the same can also be found in Ref.~\cite{Essick_2021}.

The trends in Fig.~\ref{fig:vol_q_1_imbh} at lower masses closely resemble those in Fig.~\ref{fig:vol_q_1}, as expected. For $q=1$, the $\tt{IAS-22}$ and $\tt{pycbc\_{bbh}}$ pipelines exhibit similar sensitivity volumes for $m_1^s$ below $\sim 130\, M_{\odot}$. However, beyond this threshold, the $\tt{IAS-22}$ pipeline significantly outperforms $\tt{pycbc\_{bbh}}$, with the sensitivity volume for $\tt{pycbc\_{bbh}}$ dropping to near zero beyond $m_1^s\sim 200\, M_{\odot}$ (corresponding to $M_\mathrm{tot}^{s}\sim 400\, M_{\odot}$). This decline can be attributed to the $\tt{pycbc\_{bbh}}$ pipeline's imposing a lower bound on template duration, set at $0.15$\,s during the search analysis \cite{lvc_gwtc3_o3_ab_catalog_2021,DalCanton:2017ala}. Signals from systems with $M_\mathrm{tot}^{s}\gtrsim 400\, M_{\odot}$ have durations much shorter than this threshold, leading to a loss of sensitivity. For a lower mass ratio, $q=1/4$, the $\tt{IAS-22}$ and $\tt{pycbc\_{bbh}}$ pipelines demonstrate comparable sensitivity volumes up to a relatively higher $m_1^s$ (as $M_\mathrm{tot}^{s}\sim 400\, M_{\odot}$ corresponds to $m_1^s\sim 300\, M_{\odot}$).

Having higher $M_\mathrm{tot}$ in the IMBH range can worsen detectability as the signal starts to shift out of the detector band ($f_\mathrm{min}\sim 20$ Hz). As seen in Fig.~\ref{fig:vol_q_1_imbh}, this effect leads to favoring the detectability of $q<1$ systems over equal mass-ratio systems beyond $m_1^s \gtrsim 150 M_\odot$ (as $q<1$ systems have a lower $M_\mathrm{tot}^{s}$ at given $m_1^{s}$).
% leads to another interesting effect, namely  , the asymmetric systems have a larger observable volume, which makes it even more important to use HM to detect IMBH systems.

The right panel of Fig.~\ref{fig:vol_q_1_imbh} characterizes the improvement in sensitivity volume ($\overline{VT}_{\mathrm{HM}}/\overline{VT}_{\mathrm{22}}$) upon including HM.
As both the total mass and inverse mass ratio ($1/q$) of binary mergers increase, the contribution of HM to the signal becomes more significant (see \cite[figure~1]{Wad23_TemplateBanks} and also \cite{HMeffect_ParameterSpaceDependency_PekowskyPRD2013, HMeffect_RelativeModeSignificance_HealyPRD2013,  Cap14, HMeffect_AlignedSearchImpactCalderonBustilloPRD2016, HMandPrecessionEffect_HeavySearchImpact_CalderonBustilloPRD2017, HMeffect_IMBHsearchImpact_CalderonBustilloPRD2018, Mil21, Har18, Cha22, Sha22,Zha23}). Indeed, the figure shows a clear trend of increasing fractional improvements with higher total masses for all mass ratios. 
For $q=1/4$, the inclusion of HM results in substantial sensitivity gains, with sensitivity volumes increasing by $\sim 15$--$150\%$ for primary masses in the range $\sim [100, 250]\, M_{\odot}$, corresponding to total masses of $\sim [125, 330]\, M_{\odot}$. 

It is worth noting that, beyond a certain total mass, the improvements begin to decline for all $q$. This could be due to incompleteness in $\tt{IAS}$ template banks  (only binaries till detector-frame total mass, $M_\mathrm{tot}\leq 400\, \msun$ are currently included \cite{Wad23_TemplateBanks}), and glitches dominating the background at high masses \cite{Wad23_Pipeline}. We leave a detailed analysis of this effect to a future study. We omitted the cases $q<1/4$ and $M_\mathrm{tot}^{s}>500\, \msun$, as the injection set has insufficient $N_{\mathrm{eff}}$, resulting in large uncertainties in $\overline{VT}$ and $\Delta \overline{VT}$.
% This indicates that while the $\tt{IAS-HM}$ pipeline continues to provide enhancements over $\tt{IAS-22}$ for these higher masses, the performance gains become less pronounced. We attribute this degradation at very high total masses to limitations in the construction of the rank function, which models the noise probability of obtaining a given matched-filter score \cite{Wad23_Pipeline}. At such high masses, glitches increasingly dominate the noise background, complicating signal detection. We omit the $M_{\mathrm{tot}}^{s}> 500 \, M_{\odot}$ region in the plot as the error bars significantly increase.

% Even for $q=1$, where HM contribute minimally, the $\tt{IAS-HM}$ pipeline demonstrates mild improvements over $\tt{IAS-22}$. %However, we anticipate that these sensitivity improvements will become even more pronounced for lower $q$ systems.

\section{Conclusions}
\label{sec:conclusion}

We investigated the impact of incorporating HM into the search pipelines for GW detections. We utilized injection datasets made available by the LVK collaboration for calculating the sensitivity volume-time ($\overline{VT}$) of different pipelines.
% for BBHs and IMBH binaries, spanning a wide range of mass ratios and total masses \cite{zenodoLVK}. 
% Most gravitational wave searches to date do not in their templates include higher harmonics which are predicted by general relativity.
We compared the following cases: $\tt{IAS}$ pipelines with and without including higher harmonics in the search templates; and various LVK pipelines (all of which currently only include the $(2,2)$ mode in their templates).

For high-mass and/or unequal-mass-ratio binaries, addition of HM leads to a significant improvement in sensitivity volume (Figs.~\ref{fig:vol_q_1} and~\ref{fig:vol_q_1_imbh}). The gains are particularly noteworthy in the high-mass range, enhancing $\overline{VT}$ by $\sim 15\%-150\%$ for binaries with the primary source mass, $m_1^\mathrm{s}\sim [100, 250]\, M_\odot$. For nearly equal-mass binaries, we find that including HM in our search does not lead to degradation of sensitivity. This is likely due to the fact that we use a marginalized detection statistic to upweight systems with larger observable volumes (e.g., $q\sim 1$ systems) and vice versa. We also show that the $\tt{IAS}$ pipeline with HM ($\tt{IAS-HM}$) has sensitive volume better than or comparable to that of individual LVK pipelines across the entire range of binary parameter space we probed.

% For IMBH binaries, the role of HM is even more pronounced. While the $\tt{IAS-22}$ pipeline performs comparably to the LVK $\tt{pycbc\_bbh}$ pipeline for lower primary masses, it demonstrates superior sensitivity at higher masses. The $\tt{IAS-HM}$ pipeline further amplifies this advantage, achieving gains in sensitivity volume of up to $150\%$ for lower mass ratios. 

% For low-mass BBH systems with equal masses ($q=1$) where HM (except the $44$ mode) do not contribute, we showed that the inclusion of HM does not degrade sensitivity, addressing previous concerns about increased false alarm rates or template bank complexity.
% Given that the sensitivity for IMBH binaries is inherently smaller than for BBHs, any substantial improvement in their detection volume is crucial for increasing their representation in GW catalogs.
% This is also important for getting deeper insights into populations of IMBH binaries
%These results highlight the importance of including HM in the search for IMBH binaries, where signal power and detector sensitivity are intricately balanced.

% Overall, our study underscores the value of incorporating HM in GW search pipelines, particularly for detecting more sources with high total masses (e.g., IMBHs) and unequal mass ratios. This will provide deeper insights into populations of such binaries in the universe. 
%As detectors improve in sensitivity and additional observing runs commence, leveraging HM in future searches is expected to enhance the detection rates of IMBH binaries.

We showed results corresponding to the $\tt{IAS-HM}$ pipeline$^{\ref{foot}}$ in this paper, but we expect our sensitivity results to also hold for other search pipelines if HM are introduced using the mode-by-mode filtering method. We provide the updated LVK injection summary files containing our $\tt{IAS-HM}$ and $\tt{IAS-22}$ results on Zenodo \cite{zenodoIAS}.

%%%%%%%%%%%%%%%%%%%%%%%%%%%%%%%%%%%%%%%%%%%%%%%%%%%%%%%%%%%%%%%%%%%%%%%%%%%%
\section*{Acknowledgements}
%%%%%%%%%%%%%%%%%%%%%%%%%%%%%%%%%%%%%%%%%%%%%%%%%%%%%%%%%%%%%%%%%%%%%%%%%%%%
% D. W. gratefully acknowledges support from the Friends of the Institute for Advanced Study Membership and from the W. M. Keck Foundation Fund. MZ is supported by NSF 2209991 and NSF-BSF 2207583.
% BZ is supported by NSF-BSF 2207583 and ISF

We thank Sharan Banagiri, Aaron Zimmerman, Will Farr and Maya Fishbach for useful discussions.
DW gratefully acknowledges support from the Friends of the Institute for Advanced Study Membership and the Keck foundation. 
TV acknowledges support from NSF grants 2012086 and 2309360, the Alfred P. Sloan Foundation through grant number FG-2023-20470, the BSF through award number 2022136 and the Hellman Family Faculty Fellowship. BZ is supported by the Israel Science Foundation, NSF-BSF and by a research grant from the Willner Family Leadership Institute for the Weizmann Institute of Science. MZ is supported by NSF 2209991 and NSF-BSF 2207583 and by the Nelson Center for Collaborative Research. This research was also supported in part by the National Science Foundation under Grant No. NSF PHY-1748958. We also thank ICTS-TIFR for their hospitality during the completion of a part of this work.

This research has made use of data, software and/or web tools obtained from Zenodo \cite{zenodoLVK} and the Gravitational Wave Open Science Center (\url{https://www.gw-openscience.org/}), a service of LIGO Laboratory, the LIGO Scientific Collaboration and the Virgo Collaboration. LIGO Laboratory and Advanced LIGO are funded by the United States National Science Foundation (NSF) as well as the Science and Technology Facilities Council (STFC) of the United Kingdom, the Max-Planck-Society (MPS), and the State of Niedersachsen/Germany for support of the construction of Advanced LIGO and construction and operation of the GEO600 detector. Additional support for Advanced LIGO was provided by the Australian Research Council. Virgo is funded, through the European Gravitational Observatory (EGO), by the French Centre National de Recherche Scientifique (CNRS), the Italian Istituto Nazionale di Fisica Nucleare (INFN) and the Dutch Nikhef, with contributions by institutions from Belgium, Germany, Greece, Hungary, Ireland, Japan, Monaco, Poland, Portugal, Spain.
%%%%%%%%%%%%%%%%%%%%%%%%%%%%%%%%%%%%%%%%%%%%%%%%%%%%%%%%%
%%%%%%%%%%%%%%%%%%%%%%%%%%%%%%%%%%%%%%%%%%%%%%%%%%%%%%%%%
%%%%%%%%%%%%%%%%%%%%%%%%%%%%%%%%%%%%%%%%%%%%%%%%%%%%%%%%%
%%%%%%%%%%%%%%%%%%%%%%%%%%%%%%%%%%%%%%%%%%%%%%%%%%%%%%%%%
%%%%%%%%%%%%%%%%%%%%%%%%%%%%%%%%%%%%%%%%%%%%%%%%%%%%%%%%%
%%%%%%%%%%%%%%%%%%%%%%%%%%%%%%%%%%%%%%%%%%%%%%%%%%%%%%%%%

\appendix

\section{Supplemental material}

\section{Updates to $(2, 2)$-only IAS pipeline}
The $(2, 2)$-only pipeline we used in this paper ($\tt{IAS-22}$) is the limit of the $\tt{IAS-HM}$ pipeline when higher harmonics are turned off (i.e., the SNR in the $(3,3)$ and $(4,4)$ modes is set to zero). Note that the $\tt{IAS-22}$ case in this paper is different from the previous version of the $(2, 2)$-only IAS pipeline which was used to produce the IAS O3a and O3b catalogs in Refs.~\cite{Meh23_ias_o3b}, \cite{Ols22_ias_o3a}. The $\tt{IAS-22}$ case includes all the improvements that we made for the $\tt{IAS-HM}$ search listed in Refs.~\cite{Wad23_TemplateBanks} and \cite{Wad23_Pipeline} (except, of course, the inclusion of HM). For the template banks, we use random forest regressor increase the efficiency of the banks (see Ref.~\cite[figure~1]{Wad23_TemplateBanks}). We also use band eraser to remove excess noise \cite[figure~4]{Wad23_Pipeline}. We use exactly the same astrophysical prior in the $\tt{IAS-HM}$ and $\tt{IAS-22}$ searches so one can compare the pipelines directly. Unlike the previous version of astrophysical prior used in Refs.~\cite{Meh23_ias_o3b,Ols22_ias_o3a}, our astrophysical prior includes the information of the observable volume of binaries, e.g., it upweights systems with positive effective spin ($\chieff$) compared to negative $\chieff$ \cite{Wad23_HM_Events}.

\begin{figure}[h]
    \includegraphics[width=0.5\textwidth]{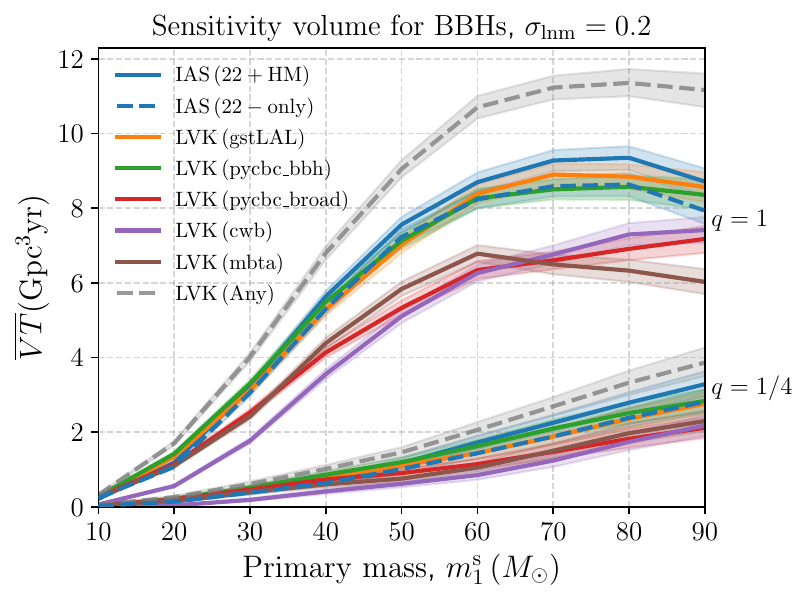}
    \vspace{-0.2cm}
    \caption{Similar to Fig.~\ref{fig:vol_q_1} in the main text, but including the special LVK-$\tt{Any}$ case in dotted and including both $q=1$ and $q=1/4$ cases. LVK-$\tt{Any}$ is an optimistic estimate which combines only the detected injections from all the
LVK pipelines (without accounting for an increase in the overall background due to the look-elsewhere effect), see the text for
details.}
    % \ajitm{Combined plot of the left and right panels of Fig.~\ref{fig:vol_q_1}, now including the results from LVK-$\tt{Any}$ case. While LVK-$\tt{Any}$ shows enhanced sensitivity compared to individual LVK pipelines and IAS pipelines, the increased rate of background triggers associated with this approach, if accounted for, could diminish the overall performance.}} 
    \label{fig:vol_any}
\end{figure}

\section{Volume sensitivity plot including $\tt{LVK-Any}$}

In Fig.~\ref{fig:vol_q_1} of the main text of the paper, we had compared the volume-time $\overline{VT}$ sensitivity of the IAS and LVK pipelines. Here, in Fig.~\ref{fig:vol_any}, we show $\overline{VT}$ for the additional case labeled as LVK-$\tt{Any}$. 
% \ajitm{In Fig.~\ref{fig:vol_any}, we revisit the $\overline{VT}$ plot from Fig.\ref{fig:vol_q_1}, now including results from an additional pipeline labeled as LVK-$\tt{Any}$.}
This case considers a trigger (or injection) in the LVK detectors as detected if any of the five LVK pipelines shown in the legend assigns it a value $p_{\mathrm{astro}}>0.5$. As expected, the combined $\overline{VT}$ is larger than that of the individual LVK pipelines, as seen in the dotted lines in Fig.~\ref{fig:vol_any}. However, this procedure will also allow for additional background (as a statistic that maximizes over $p_\mathrm{astro}$ among pipelines would also produce a larger number of background triggers above a $p_\mathrm{astro}$ value than the individual pipelines). See Refs.~\cite{Banagiri:2023ztt, Bis12_Combining_Pipelines} for further discussion, where they also propose statistically consistent ways to combine pipelines by reweighting the individual $p_\mathrm{astro}$ values.

% This outperforms the individual LVK pipelines and the IAS pipelines, particularly for $q=1$ binaries. However, maximizing over $p_\mathrm{astro}$ values from individual pipelines inherently increases the rate of background triggers exceeding the $p_\mathrm{astro}$ threshold. Accounting for this effect would reduce the effective performance of LVK-$\tt{Any}$ \cite{Banagiri:2023ztt, Bis12_Combining_Pipelines}.

\bibliographystyle{apsrev4-1-etal}
\bibliography{main}

\begin{thebibliography}{75}%
\makeatletter
\providecommand \@ifxundefined [1]{%
 \@ifx{#1\undefined}
}%
\providecommand \@ifnum [1]{%
 \ifnum #1\expandafter \@firstoftwo
 \else \expandafter \@secondoftwo
 \fi
}%
\providecommand \@ifx [1]{%
 \ifx #1\expandafter \@firstoftwo
 \else \expandafter \@secondoftwo
 \fi
}%
\providecommand \natexlab [1]{#1}%
\providecommand \enquote  [1]{``#1''}%
\providecommand \bibnamefont  [1]{#1}%
\providecommand \bibfnamefont [1]{#1}%
\providecommand \citenamefont [1]{#1}%
\providecommand \href@noop [0]{\@secondoftwo}%
\providecommand \href [0]{\begingroup \@sanitize@url \@href}%
\providecommand \@href[1]{\@@startlink{#1}\@@href}%
\providecommand \@@href[1]{\endgroup#1\@@endlink}%
\providecommand \@sanitize@url [0]{\catcode `\\12\catcode `\$12\catcode `\&12\catcode `\#12\catcode `\^12\catcode `\_12\catcode `\%12\relax}%
\providecommand \@@startlink[1]{}%
\providecommand \@@endlink[0]{}%
\providecommand \url  [0]{\begingroup\@sanitize@url \@url }%
\providecommand \@url [1]{\endgroup\@href {#1}{\urlprefix }}%
\providecommand \urlprefix  [0]{URL }%
\providecommand \Eprint [0]{\href }%
\providecommand \doibase [0]{http://dx.doi.org/}%
\providecommand \selectlanguage [0]{\@gobble}%
\providecommand \bibinfo  [0]{\@secondoftwo}%
\providecommand \bibfield  [0]{\@secondoftwo}%
\providecommand \translation [1]{[#1]}%
\providecommand \BibitemOpen [0]{}%
\providecommand \bibitemStop [0]{}%
\providecommand \bibitemNoStop [0]{.\EOS\space}%
\providecommand \EOS [0]{\spacefactor3000\relax}%
\providecommand \BibitemShut  [1]{\csname bibitem#1\endcsname}%
\let\auto@bib@innerbib\@empty
%</preamble>
\bibitem [{\citenamefont {{Wadekar}}\ \emph {et~al.}(2024)\citenamefont {{Wadekar}}, \citenamefont {{Venumadhav}}, \citenamefont {{Roulet}}, \citenamefont {{Mehta}}, \citenamefont {{Zackay}}, \citenamefont {{Mushkin}},\ and\ \citenamefont {{Zaldarriaga}}}]{Wad23_Pipeline}%
  \BibitemOpen
  \bibfield  {author} {\bibinfo {author} {\bibfnamefont {D.}~\bibnamefont {{Wadekar}}}, \bibinfo {author} {\bibfnamefont {T.}~\bibnamefont {{Venumadhav}}}, \bibinfo {author} {\bibfnamefont {J.}~\bibnamefont {{Roulet}}}, \bibinfo {author} {\bibfnamefont {A.~K.}\ \bibnamefont {{Mehta}}}, \bibinfo {author} {\bibfnamefont {B.}~\bibnamefont {{Zackay}}}, \bibinfo {author} {\bibfnamefont {J.}~\bibnamefont {{Mushkin}}}, \ and\ \bibinfo {author} {\bibfnamefont {M.}~\bibnamefont {{Zaldarriaga}}},\ }\href {\doibase 10.1103/PhysRevD.110.044063} {\bibfield  {journal} {\bibinfo  {journal} {\prd}\ }\textbf {\bibinfo {volume} {110}},\ \bibinfo {eid} {044063} (\bibinfo {year} {2024})},\ \Eprint {http://arxiv.org/abs/2405.17400} {arXiv:2405.17400 [gr-qc]} \BibitemShut {NoStop}%
\bibitem [{\citenamefont {Aasi}\ \emph {et~al.}(2015)\citenamefont {Aasi} \emph {et~al.}}]{LIGOScientific:2014pky}%
  \BibitemOpen
  \bibfield  {author} {\bibinfo {author} {\bibfnamefont {J.}~\bibnamefont {Aasi}} \emph {et~al.} (\bibinfo {collaboration} {LIGO Scientific}),\ }\href {\doibase 10.1088/0264-9381/32/7/074001} {\bibfield  {journal} {\bibinfo  {journal} {Class. Quant. Grav.}\ }\textbf {\bibinfo {volume} {32}},\ \bibinfo {pages} {074001} (\bibinfo {year} {2015})},\ \Eprint {http://arxiv.org/abs/1411.4547} {arXiv:1411.4547 [gr-qc]} \BibitemShut {NoStop}%
\bibitem [{\citenamefont {Acernese}\ \emph {et~al.}(2015)\citenamefont {Acernese} \emph {et~al.}}]{VIRGO:2014yos}%
  \BibitemOpen
  \bibfield  {author} {\bibinfo {author} {\bibfnamefont {F.}~\bibnamefont {Acernese}} \emph {et~al.} (\bibinfo {collaboration} {VIRGO}),\ }\href {\doibase 10.1088/0264-9381/32/2/024001} {\bibfield  {journal} {\bibinfo  {journal} {Class. Quant. Grav.}\ }\textbf {\bibinfo {volume} {32}},\ \bibinfo {pages} {024001} (\bibinfo {year} {2015})},\ \Eprint {http://arxiv.org/abs/1408.3978} {arXiv:1408.3978 [gr-qc]} \BibitemShut {NoStop}%
\bibitem [{\citenamefont {Akutsu}\ \emph {et~al.}(2021)\citenamefont {Akutsu} \emph {et~al.}}]{KAGRA:2020tym}%
  \BibitemOpen
  \bibfield  {author} {\bibinfo {author} {\bibfnamefont {T.}~\bibnamefont {Akutsu}} \emph {et~al.} (\bibinfo {collaboration} {KAGRA}),\ }\href {\doibase 10.1093/ptep/ptaa125} {\bibfield  {journal} {\bibinfo  {journal} {PTEP}\ }\textbf {\bibinfo {volume} {2021}},\ \bibinfo {pages} {05A101} (\bibinfo {year} {2021})},\ \Eprint {http://arxiv.org/abs/2005.05574} {arXiv:2005.05574 [physics.ins-det]} \BibitemShut {NoStop}%
\bibitem [{\citenamefont {Cannon}\ \emph {et~al.}(2021)\citenamefont {Cannon}, \citenamefont {Caudill}, \citenamefont {Chan}, \citenamefont {Cousins}, \citenamefont {Creighton}, \citenamefont {Ewing}, \citenamefont {Fong}, \citenamefont {Godwin}, \citenamefont {Hanna}, \citenamefont {Hooper}, \citenamefont {Huxford}, \citenamefont {Magee}, \citenamefont {Meacher}, \citenamefont {Messick}, \citenamefont {Morisaki}, \citenamefont {Mukherjee}, \citenamefont {Ohta}, \citenamefont {Pace}, \citenamefont {Privitera}, \citenamefont {{de Ruiter}}, \citenamefont {Sachdev}, \citenamefont {Singer}, \citenamefont {Singh}, \citenamefont {Tapia}, \citenamefont {Tsukada}, \citenamefont {Tsuna}, \citenamefont {Tsutsui}, \citenamefont {Ueno}, \citenamefont {Viets}, \citenamefont {Wade},\ and\ \citenamefont {Wade}}]{CANNON2021100680}%
  \BibitemOpen
  \bibfield  {author} {\bibinfo {author} {\bibfnamefont {K.}~\bibnamefont {Cannon}}, \bibinfo {author} {\bibfnamefont {S.}~\bibnamefont {Caudill}}, \bibinfo {author} {\bibfnamefont {C.}~\bibnamefont {Chan}}, \bibinfo {author} {\bibfnamefont {B.}~\bibnamefont {Cousins}},  \emph {et~al.},\ }\href {\doibase https://doi.org/10.1016/j.softx.2021.100680} {\bibfield  {journal} {\bibinfo  {journal} {SoftwareX}\ }\textbf {\bibinfo {volume} {14}},\ \bibinfo {pages} {100680} (\bibinfo {year} {2021})}\BibitemShut {NoStop}%
\bibitem [{\citenamefont {Aubin}\ \emph {et~al.}(2021)\citenamefont {Aubin} \emph {et~al.}}]{Aubin:2020goo}%
  \BibitemOpen
  \bibfield  {author} {\bibinfo {author} {\bibfnamefont {F.}~\bibnamefont {Aubin}} \emph {et~al.},\ }\href {\doibase 10.1088/1361-6382/abe913} {\bibfield  {journal} {\bibinfo  {journal} {Class. Quant. Grav.}\ }\textbf {\bibinfo {volume} {38}},\ \bibinfo {pages} {095004} (\bibinfo {year} {2021})},\ \Eprint {http://arxiv.org/abs/2012.11512} {arXiv:2012.11512 [gr-qc]} \BibitemShut {NoStop}%
\bibitem [{\citenamefont {Klimenko}\ \emph {et~al.}(2016)\citenamefont {Klimenko} \emph {et~al.}}]{Klimenko:2015ypf}%
  \BibitemOpen
  \bibfield  {author} {\bibinfo {author} {\bibfnamefont {S.}~\bibnamefont {Klimenko}} \emph {et~al.},\ }\href {\doibase 10.1103/PhysRevD.93.042004} {\bibfield  {journal} {\bibinfo  {journal} {Phys. Rev. D}\ }\textbf {\bibinfo {volume} {93}},\ \bibinfo {pages} {042004} (\bibinfo {year} {2016})},\ \Eprint {http://arxiv.org/abs/1511.05999} {arXiv:1511.05999 [gr-qc]} \BibitemShut {NoStop}%
\bibitem [{\citenamefont {Usman}\ \emph {et~al.}(2016)\citenamefont {Usman} \emph {et~al.}}]{Usman:2015kfa}%
  \BibitemOpen
  \bibfield  {author} {\bibinfo {author} {\bibfnamefont {S.~A.}\ \bibnamefont {Usman}} \emph {et~al.},\ }\href {\doibase 10.1088/0264-9381/33/21/215004} {\bibfield  {journal} {\bibinfo  {journal} {Class. Quant. Grav.}\ }\textbf {\bibinfo {volume} {33}},\ \bibinfo {pages} {215004} (\bibinfo {year} {2016})},\ \Eprint {http://arxiv.org/abs/1508.02357} {arXiv:1508.02357 [gr-qc]} \BibitemShut {NoStop}%
\bibitem [{\citenamefont {Venumadhav}\ \emph {et~al.}(2019)\citenamefont {Venumadhav}, \citenamefont {Zackay}, \citenamefont {Roulet}, \citenamefont {Dai},\ and\ \citenamefont {Zaldarriaga}}]{ias_pipeline_o1_catalog_new_search_prd2019}%
  \BibitemOpen
  \bibfield  {author} {\bibinfo {author} {\bibfnamefont {T.}~\bibnamefont {Venumadhav}}, \bibinfo {author} {\bibfnamefont {B.}~\bibnamefont {Zackay}}, \bibinfo {author} {\bibfnamefont {J.}~\bibnamefont {Roulet}}, \bibinfo {author} {\bibfnamefont {L.}~\bibnamefont {Dai}}, \ and\ \bibinfo {author} {\bibfnamefont {M.}~\bibnamefont {Zaldarriaga}},\ }\href {\doibase 10.1103/PhysRevD.100.023011} {\bibfield  {journal} {\bibinfo  {journal} {Phys. Rev.}\ }\textbf {\bibinfo {volume} {D100}},\ \bibinfo {pages} {023011} (\bibinfo {year} {2019})},\ \Eprint {http://arxiv.org/abs/1902.10341} {arXiv:1902.10341 [astro-ph.IM]} \BibitemShut {NoStop}%
%%CITATION = ARXIV:1902.10341;%%
\bibitem [{\citenamefont {{Venumadhav}}\ \emph {et~al.}(2020)\citenamefont {{Venumadhav}}, \citenamefont {{Zackay}}, \citenamefont {{Roulet}}, \citenamefont {{Dai}},\ and\ \citenamefont {{Zaldarriaga}}}]{ias_o2_pipeline_new_events_prd2020}%
  \BibitemOpen
  \bibfield  {author} {\bibinfo {author} {\bibfnamefont {T.}~\bibnamefont {{Venumadhav}}}, \bibinfo {author} {\bibfnamefont {B.}~\bibnamefont {{Zackay}}}, \bibinfo {author} {\bibfnamefont {J.}~\bibnamefont {{Roulet}}}, \bibinfo {author} {\bibfnamefont {L.}~\bibnamefont {{Dai}}}, \ and\ \bibinfo {author} {\bibfnamefont {M.}~\bibnamefont {{Zaldarriaga}}},\ }\href {\doibase 10.1103/PhysRevD.101.083030} {\bibfield  {journal} {\bibinfo  {journal} {\prd}\ }\textbf {\bibinfo {volume} {101}},\ \bibinfo {eid} {083030} (\bibinfo {year} {2020})},\ \Eprint {http://arxiv.org/abs/1904.07214} {arXiv:1904.07214 [astro-ph.HE]} \BibitemShut {NoStop}%
\bibitem [{\citenamefont {Olsen}\ \emph {et~al.}(2022)\citenamefont {Olsen}, \citenamefont {Venumadhav}, \citenamefont {Mushkin}, \citenamefont {Roulet}, \citenamefont {Zackay},\ and\ \citenamefont {Zaldarriaga}}]{Ols22_ias_o3a}%
  \BibitemOpen
  \bibfield  {author} {\bibinfo {author} {\bibfnamefont {S.}~\bibnamefont {Olsen}}, \bibinfo {author} {\bibfnamefont {T.}~\bibnamefont {Venumadhav}}, \bibinfo {author} {\bibfnamefont {J.}~\bibnamefont {Mushkin}}, \bibinfo {author} {\bibfnamefont {J.}~\bibnamefont {Roulet}}, \bibinfo {author} {\bibfnamefont {B.}~\bibnamefont {Zackay}}, \ and\ \bibinfo {author} {\bibfnamefont {M.}~\bibnamefont {Zaldarriaga}},\ }\href {\doibase 10.1103/PhysRevD.106.043009} {\bibfield  {journal} {\bibinfo  {journal} {Phys. Rev. D}\ }\textbf {\bibinfo {volume} {106}},\ \bibinfo {pages} {043009} (\bibinfo {year} {2022})}\BibitemShut {NoStop}%
\bibitem [{\citenamefont {Nitz}\ \emph {et~al.}(2019)\citenamefont {Nitz}, \citenamefont {Capano}, \citenamefont {Nielsen}, \citenamefont {Reyes}, \citenamefont {White}, \citenamefont {Brown},\ and\ \citenamefont {Krishnan}}]{NitzCatalog_1-OGC_o1_2018}%
  \BibitemOpen
  \bibfield  {author} {\bibinfo {author} {\bibfnamefont {A.~H.}\ \bibnamefont {Nitz}}, \bibinfo {author} {\bibfnamefont {C.}~\bibnamefont {Capano}}, \bibinfo {author} {\bibfnamefont {A.~B.}\ \bibnamefont {Nielsen}}, \bibinfo {author} {\bibfnamefont {S.}~\bibnamefont {Reyes}}, \bibinfo {author} {\bibfnamefont {R.}~\bibnamefont {White}}, \bibinfo {author} {\bibfnamefont {D.~A.}\ \bibnamefont {Brown}}, \ and\ \bibinfo {author} {\bibfnamefont {B.}~\bibnamefont {Krishnan}},\ }\href {\doibase 10.3847/1538-4357/ab0108} {\bibfield  {journal} {\bibinfo  {journal} {The Astrophysical Journal}\ }\textbf {\bibinfo {volume} {872}},\ \bibinfo {pages} {195} (\bibinfo {year} {2019})}\BibitemShut {NoStop}%
\bibitem [{\citenamefont {Nitz}\ \emph {et~al.}(2020)\citenamefont {Nitz}, \citenamefont {Dent}, \citenamefont {Davies}, \citenamefont {Kumar}, \citenamefont {Capano}, \citenamefont {Harry}, \citenamefont {Mozzon}, \citenamefont {Nuttall}, \citenamefont {Lundgren},\ and\ \citenamefont {T\'apai}}]{NitzCatalog_2-OGC_o2_2020}%
  \BibitemOpen
  \bibfield  {author} {\bibinfo {author} {\bibfnamefont {A.~H.}\ \bibnamefont {Nitz}}, \bibinfo {author} {\bibfnamefont {T.}~\bibnamefont {Dent}}, \bibinfo {author} {\bibfnamefont {G.~S.}\ \bibnamefont {Davies}}, \bibinfo {author} {\bibfnamefont {S.}~\bibnamefont {Kumar}},  \emph {et~al.},\ }\href {\doibase 10.3847/1538-4357/ab733f} {\bibfield  {journal} {\bibinfo  {journal} {Astrophys. J.}\ }\textbf {\bibinfo {volume} {891}},\ \bibinfo {pages} {123} (\bibinfo {year} {2020})},\ \Eprint {http://arxiv.org/abs/1910.05331} {arXiv:1910.05331 [astro-ph.HE]} \BibitemShut {NoStop}%
\bibitem [{\citenamefont {{Nitz}}\ \emph {et~al.}(2021)\citenamefont {{Nitz}}, \citenamefont {{Capano}}, \citenamefont {{Kumar}}, \citenamefont {{Wang}}, \citenamefont {{Kastha}}, \citenamefont {{Sch{\"a}fer}}, \citenamefont {{Dhurkunde}},\ and\ \citenamefont {{Cabero}}}]{nitz_o3a_3ogc_catalog_2021}%
  \BibitemOpen
  \bibfield  {author} {\bibinfo {author} {\bibfnamefont {A.~H.}\ \bibnamefont {{Nitz}}}, \bibinfo {author} {\bibfnamefont {C.~D.}\ \bibnamefont {{Capano}}}, \bibinfo {author} {\bibfnamefont {S.}~\bibnamefont {{Kumar}}}, \bibinfo {author} {\bibfnamefont {Y.-F.}\ \bibnamefont {{Wang}}},  \emph {et~al.},\ }\href {https://ui.adsabs.harvard.edu/abs/2021arXiv210509151N} {\bibfield  {journal} {\bibinfo  {journal} {arXiv e-prints}\ ,\ \bibinfo {eid} {arXiv:2105.09151}} (\bibinfo {year} {2021})},\ \Eprint {http://arxiv.org/abs/2105.09151} {arXiv:2105.09151 [astro-ph.HE]} \BibitemShut {NoStop}%
\bibitem [{\citenamefont {Nitz}\ \emph {et~al.}(2023)\citenamefont {Nitz}, \citenamefont {Kumar}, \citenamefont {Wang}, \citenamefont {Kastha}, \citenamefont {Wu}, \citenamefont {Sch\"{a}fer}, \citenamefont {Dhurkunde},\ and\ \citenamefont {Capano}}]{nitz_4ogc_o3_ab_catalog_2021}%
  \BibitemOpen
  \bibfield  {author} {\bibinfo {author} {\bibfnamefont {A.~H.}\ \bibnamefont {Nitz}}, \bibinfo {author} {\bibfnamefont {S.}~\bibnamefont {Kumar}}, \bibinfo {author} {\bibfnamefont {Y.-F.}\ \bibnamefont {Wang}}, \bibinfo {author} {\bibfnamefont {S.}~\bibnamefont {Kastha}},  \emph {et~al.},\ }\href {\doibase 10.3847/1538-4357/aca591} {\bibfield  {journal} {\bibinfo  {journal} {The Astrophysical Journal}\ }\textbf {\bibinfo {volume} {946}},\ \bibinfo {pages} {59} (\bibinfo {year} {2023})}\BibitemShut {NoStop}%
\bibitem [{\citenamefont {{Chia}}\ \emph {et~al.}(2024)\citenamefont {{Chia}}, \citenamefont {{Edwards}}, \citenamefont {{Wadekar}}, \citenamefont {{Zimmerman}}, \citenamefont {{Olsen}}, \citenamefont {{Roulet}}, \citenamefont {{Venumadhav}}, \citenamefont {{Zackay}},\ and\ \citenamefont {{Zaldarriaga}}}]{Chi23}%
  \BibitemOpen
  \bibfield  {author} {\bibinfo {author} {\bibfnamefont {H.~S.}\ \bibnamefont {{Chia}}}, \bibinfo {author} {\bibfnamefont {T.~D.~P.}\ \bibnamefont {{Edwards}}}, \bibinfo {author} {\bibfnamefont {D.}~\bibnamefont {{Wadekar}}}, \bibinfo {author} {\bibfnamefont {A.}~\bibnamefont {{Zimmerman}}},  \emph {et~al.},\ }\href {\doibase 10.1103/PhysRevD.110.063007} {\bibfield  {journal} {\bibinfo  {journal} {\prd}\ }\textbf {\bibinfo {volume} {110}},\ \bibinfo {eid} {063007} (\bibinfo {year} {2024})},\ \Eprint {http://arxiv.org/abs/2306.00050} {arXiv:2306.00050 [gr-qc]} \BibitemShut {NoStop}%
\bibitem [{\citenamefont {{Mehta}}\ \emph {et~al.}(2023)\citenamefont {{Mehta}}, \citenamefont {{Olsen}}, \citenamefont {{Wadekar}}, \citenamefont {{Roulet}}, \citenamefont {{Venumadhav}}, \citenamefont {{Mushkin}}, \citenamefont {{Zackay}},\ and\ \citenamefont {{Zaldarriaga}}}]{Meh23_ias_o3b}%
  \BibitemOpen
  \bibfield  {author} {\bibinfo {author} {\bibfnamefont {A.~K.}\ \bibnamefont {{Mehta}}}, \bibinfo {author} {\bibfnamefont {S.}~\bibnamefont {{Olsen}}}, \bibinfo {author} {\bibfnamefont {D.}~\bibnamefont {{Wadekar}}}, \bibinfo {author} {\bibfnamefont {J.}~\bibnamefont {{Roulet}}},  \emph {et~al.},\ }\href {\doibase 10.48550/arXiv.2311.06061} {\bibfield  {journal} {\bibinfo  {journal} {arXiv e-prints}\ ,\ \bibinfo {eid} {arXiv:2311.06061}} (\bibinfo {year} {2023})},\ \Eprint {http://arxiv.org/abs/2311.06061} {arXiv:2311.06061 [gr-qc]} \BibitemShut {NoStop}%
\bibitem [{\citenamefont {{Koloniari}}\ \emph {et~al.}(2024)\citenamefont {{Koloniari}}, \citenamefont {{Koursoumpa}}, \citenamefont {{Nousi}}, \citenamefont {{Lampropoulos}}, \citenamefont {{Passalis}}, \citenamefont {{Tefas}},\ and\ \citenamefont {{Stergioulas}}}]{Kol24_Ares_ML_Search}%
  \BibitemOpen
  \bibfield  {author} {\bibinfo {author} {\bibfnamefont {A.~E.}\ \bibnamefont {{Koloniari}}}, \bibinfo {author} {\bibfnamefont {E.~C.}\ \bibnamefont {{Koursoumpa}}}, \bibinfo {author} {\bibfnamefont {P.}~\bibnamefont {{Nousi}}}, \bibinfo {author} {\bibfnamefont {P.}~\bibnamefont {{Lampropoulos}}}, \bibinfo {author} {\bibfnamefont {N.}~\bibnamefont {{Passalis}}}, \bibinfo {author} {\bibfnamefont {A.}~\bibnamefont {{Tefas}}}, \ and\ \bibinfo {author} {\bibfnamefont {N.}~\bibnamefont {{Stergioulas}}},\ }\href@noop {} {\bibfield  {journal} {\bibinfo  {journal} {arXiv e-prints}\ ,\ \bibinfo {eid} {arXiv:2407.07820}} (\bibinfo {year} {2024})},\ \Eprint {http://arxiv.org/abs/2407.07820} {arXiv:2407.07820 [gr-qc]} \BibitemShut {NoStop}%
\bibitem [{\citenamefont {Abbott}\ \emph {et~al.}(2023{\natexlab{a}})\citenamefont {Abbott} \emph {et~al.}}]{lvc_gwtc3_o3_ab_catalog_2021}%
  \BibitemOpen
  \bibfield  {author} {\bibinfo {author} {\bibfnamefont {R.}~\bibnamefont {Abbott}} \emph {et~al.} (\bibinfo {collaboration} {LIGO Scientific Collaboration, Virgo Collaboration, and KAGRA Collaboration}),\ }\href {\doibase 10.1103/PhysRevX.13.041039} {\bibfield  {journal} {\bibinfo  {journal} {Phys. Rev. X}\ }\textbf {\bibinfo {volume} {13}},\ \bibinfo {pages} {041039} (\bibinfo {year} {2023}{\natexlab{a}})}\BibitemShut {NoStop}%
\bibitem [{\citenamefont {{Wadekar}}\ \emph {et~al.}(2023{\natexlab{a}})\citenamefont {{Wadekar}}, \citenamefont {{Roulet}}, \citenamefont {{Venumadhav}}, \citenamefont {{Mehta}}, \citenamefont {{Zackay}}, \citenamefont {{Mushkin}}, \citenamefont {{Olsen}},\ and\ \citenamefont {{Zaldarriaga}}}]{Wad23_HM_Events}%
  \BibitemOpen
  \bibfield  {author} {\bibinfo {author} {\bibfnamefont {D.}~\bibnamefont {{Wadekar}}}, \bibinfo {author} {\bibfnamefont {J.}~\bibnamefont {{Roulet}}}, \bibinfo {author} {\bibfnamefont {T.}~\bibnamefont {{Venumadhav}}}, \bibinfo {author} {\bibfnamefont {A.~K.}\ \bibnamefont {{Mehta}}},  \emph {et~al.},\ }\href@noop {} {\bibfield  {journal} {\bibinfo  {journal} {arXiv e-prints}\ ,\ \bibinfo {eid} {arXiv:2312.06631}} (\bibinfo {year} {2023}{\natexlab{a}})},\ \Eprint {http://arxiv.org/abs/2312.06631} {arXiv:2312.06631 [gr-qc]} \BibitemShut {NoStop}%
\bibitem [{\citenamefont {Abbott}\ \emph {et~al.}(2023{\natexlab{b}})\citenamefont {Abbott} \emph {et~al.}}]{LVKO3bpopulation}%
  \BibitemOpen
  \bibfield  {author} {\bibinfo {author} {\bibfnamefont {R.}~\bibnamefont {Abbott}} \emph {et~al.} (\bibinfo {collaboration} {KAGRA, VIRGO, LIGO Scientific}),\ }\href {\doibase 10.1103/PhysRevX.13.011048} {\bibfield  {journal} {\bibinfo  {journal} {Phys. Rev. X}\ }\textbf {\bibinfo {volume} {13}},\ \bibinfo {pages} {011048} (\bibinfo {year} {2023}{\natexlab{b}})},\ \Eprint {http://arxiv.org/abs/2111.03634} {arXiv:2111.03634 [astro-ph.HE]} \BibitemShut {NoStop}%
\bibitem [{\citenamefont {Roulet}\ \emph {et~al.}(2021)\citenamefont {Roulet}, \citenamefont {Chia}, \citenamefont {Olsen}, \citenamefont {Dai}, \citenamefont {Venumadhav}, \citenamefont {Zackay},\ and\ \citenamefont {Zaldarriaga}}]{ias_o3a_population_analysis_prd2021roulet}%
  \BibitemOpen
  \bibfield  {author} {\bibinfo {author} {\bibfnamefont {J.}~\bibnamefont {Roulet}}, \bibinfo {author} {\bibfnamefont {H.~S.}\ \bibnamefont {Chia}}, \bibinfo {author} {\bibfnamefont {S.}~\bibnamefont {Olsen}}, \bibinfo {author} {\bibfnamefont {L.}~\bibnamefont {Dai}}, \bibinfo {author} {\bibfnamefont {T.}~\bibnamefont {Venumadhav}}, \bibinfo {author} {\bibfnamefont {B.}~\bibnamefont {Zackay}}, \ and\ \bibinfo {author} {\bibfnamefont {M.}~\bibnamefont {Zaldarriaga}},\ }\href {\doibase 10.1103/PhysRevD.104.083010} {\bibfield  {journal} {\bibinfo  {journal} {Phys. Rev. D}\ }\textbf {\bibinfo {volume} {104}},\ \bibinfo {pages} {083010} (\bibinfo {year} {2021})}\BibitemShut {NoStop}%
\bibitem [{\citenamefont {Abbott}\ \emph {et~al.}(2021)\citenamefont {Abbott} \emph {et~al.}}]{LIGOScientific:2021sio}%
  \BibitemOpen
  \bibfield  {author} {\bibinfo {author} {\bibfnamefont {R.}~\bibnamefont {Abbott}} \emph {et~al.} (\bibinfo {collaboration} {LIGO Scientific, VIRGO, KAGRA}),\ }\href@noop {} {\  (\bibinfo {year} {2021})},\ \Eprint {http://arxiv.org/abs/2112.06861} {arXiv:2112.06861 [gr-qc]} \BibitemShut {NoStop}%
\bibitem [{\citenamefont {{Wadekar}}\ \emph {et~al.}(2023{\natexlab{b}})\citenamefont {{Wadekar}}, \citenamefont {{Venumadhav}}, \citenamefont {{Mehta}}, \citenamefont {{Roulet}}, \citenamefont {{Olsen}}, \citenamefont {{Mushkin}}, \citenamefont {{Zackay}},\ and\ \citenamefont {{Zaldarriaga}}}]{Wad23_TemplateBanks}%
  \BibitemOpen
  \bibfield  {author} {\bibinfo {author} {\bibfnamefont {D.}~\bibnamefont {{Wadekar}}}, \bibinfo {author} {\bibfnamefont {T.}~\bibnamefont {{Venumadhav}}}, \bibinfo {author} {\bibfnamefont {A.~K.}\ \bibnamefont {{Mehta}}}, \bibinfo {author} {\bibfnamefont {J.}~\bibnamefont {{Roulet}}},  \emph {et~al.},\ }\href@noop {} {\bibfield  {journal} {\bibinfo  {journal} {arXiv e-prints}\ ,\ \bibinfo {eid} {arXiv:2310.15233}} (\bibinfo {year} {2023}{\natexlab{b}})},\ \Eprint {http://arxiv.org/abs/2310.15233} {arXiv:2310.15233 [gr-qc]} \BibitemShut {NoStop}%
\bibitem [{\citenamefont {Pekowsky}\ \emph {et~al.}(2013)\citenamefont {Pekowsky}, \citenamefont {Healy}, \citenamefont {Shoemaker},\ and\ \citenamefont {Laguna}}]{HMeffect_ParameterSpaceDependency_PekowskyPRD2013}%
  \BibitemOpen
  \bibfield  {author} {\bibinfo {author} {\bibfnamefont {L.}~\bibnamefont {Pekowsky}}, \bibinfo {author} {\bibfnamefont {J.}~\bibnamefont {Healy}}, \bibinfo {author} {\bibfnamefont {D.}~\bibnamefont {Shoemaker}}, \ and\ \bibinfo {author} {\bibfnamefont {P.}~\bibnamefont {Laguna}},\ }\href {\doibase 10.1103/physrevd.87.084008} {\bibfield  {journal} {\bibinfo  {journal} {Physical Review D}\ }\textbf {\bibinfo {volume} {87}} (\bibinfo {year} {2013}),\ 10.1103/physrevd.87.084008}\BibitemShut {NoStop}%
\bibitem [{\citenamefont {Healy}\ \emph {et~al.}(2013)\citenamefont {Healy}, \citenamefont {Laguna}, \citenamefont {Pekowsky},\ and\ \citenamefont {Shoemaker}}]{HMeffect_RelativeModeSignificance_HealyPRD2013}%
  \BibitemOpen
  \bibfield  {author} {\bibinfo {author} {\bibfnamefont {J.}~\bibnamefont {Healy}}, \bibinfo {author} {\bibfnamefont {P.}~\bibnamefont {Laguna}}, \bibinfo {author} {\bibfnamefont {L.}~\bibnamefont {Pekowsky}}, \ and\ \bibinfo {author} {\bibfnamefont {D.}~\bibnamefont {Shoemaker}},\ }\href {\doibase 10.1103/physrevd.88.024034} {\bibfield  {journal} {\bibinfo  {journal} {Physical Review D}\ }\textbf {\bibinfo {volume} {88}} (\bibinfo {year} {2013}),\ 10.1103/physrevd.88.024034}\BibitemShut {NoStop}%
\bibitem [{\citenamefont {Capano}\ \emph {et~al.}(2014)\citenamefont {Capano}, \citenamefont {Pan},\ and\ \citenamefont {Buonanno}}]{Cap14}%
  \BibitemOpen
  \bibfield  {author} {\bibinfo {author} {\bibfnamefont {C.}~\bibnamefont {Capano}}, \bibinfo {author} {\bibfnamefont {Y.}~\bibnamefont {Pan}}, \ and\ \bibinfo {author} {\bibfnamefont {A.}~\bibnamefont {Buonanno}},\ }\href {\doibase 10.1103/physrevd.89.102003} {\bibfield  {journal} {\bibinfo  {journal} {Physical Review D}\ }\textbf {\bibinfo {volume} {89}} (\bibinfo {year} {2014}),\ 10.1103/physrevd.89.102003}\BibitemShut {NoStop}%
\bibitem [{\citenamefont {Bustillo}\ \emph {et~al.}(2016)\citenamefont {Bustillo}, \citenamefont {Husa}, \citenamefont {Sintes},\ and\ \citenamefont {Pürrer}}]{HMeffect_AlignedSearchImpactCalderonBustilloPRD2016}%
  \BibitemOpen
  \bibfield  {author} {\bibinfo {author} {\bibfnamefont {J.~C.}\ \bibnamefont {Bustillo}}, \bibinfo {author} {\bibfnamefont {S.}~\bibnamefont {Husa}}, \bibinfo {author} {\bibfnamefont {A.~M.}\ \bibnamefont {Sintes}}, \ and\ \bibinfo {author} {\bibfnamefont {M.}~\bibnamefont {Pürrer}},\ }\href {\doibase 10.1103/physrevd.93.084019} {\bibfield  {journal} {\bibinfo  {journal} {Physical Review D}\ }\textbf {\bibinfo {volume} {93}} (\bibinfo {year} {2016}),\ 10.1103/physrevd.93.084019}\BibitemShut {NoStop}%
\bibitem [{\citenamefont {Bustillo}\ \emph {et~al.}(2017)\citenamefont {Bustillo}, \citenamefont {Laguna},\ and\ \citenamefont {Shoemaker}}]{HMandPrecessionEffect_HeavySearchImpact_CalderonBustilloPRD2017}%
  \BibitemOpen
  \bibfield  {author} {\bibinfo {author} {\bibfnamefont {J.~C.}\ \bibnamefont {Bustillo}}, \bibinfo {author} {\bibfnamefont {P.}~\bibnamefont {Laguna}}, \ and\ \bibinfo {author} {\bibfnamefont {D.}~\bibnamefont {Shoemaker}},\ }\href {\doibase 10.1103/physrevd.95.104038} {\bibfield  {journal} {\bibinfo  {journal} {Physical Review D}\ }\textbf {\bibinfo {volume} {95}} (\bibinfo {year} {2017}),\ 10.1103/physrevd.95.104038}\BibitemShut {NoStop}%
\bibitem [{\citenamefont {Bustillo}\ \emph {et~al.}(2018)\citenamefont {Bustillo}, \citenamefont {Salemi}, \citenamefont {Canton},\ and\ \citenamefont {Jani}}]{HMeffect_IMBHsearchImpact_CalderonBustilloPRD2018}%
  \BibitemOpen
  \bibfield  {author} {\bibinfo {author} {\bibfnamefont {J.~C.}\ \bibnamefont {Bustillo}}, \bibinfo {author} {\bibfnamefont {F.}~\bibnamefont {Salemi}}, \bibinfo {author} {\bibfnamefont {T.~D.}\ \bibnamefont {Canton}}, \ and\ \bibinfo {author} {\bibfnamefont {K.~P.}\ \bibnamefont {Jani}},\ }\href {\doibase 10.1103/physrevd.97.024016} {\bibfield  {journal} {\bibinfo  {journal} {Physical Review D}\ }\textbf {\bibinfo {volume} {97}} (\bibinfo {year} {2018}),\ 10.1103/physrevd.97.024016}\BibitemShut {NoStop}%
\bibitem [{\citenamefont {{Mills}}\ and\ \citenamefont {{Fairhurst}}(2021)}]{Mil21}%
  \BibitemOpen
  \bibfield  {author} {\bibinfo {author} {\bibfnamefont {C.}~\bibnamefont {{Mills}}}\ and\ \bibinfo {author} {\bibfnamefont {S.}~\bibnamefont {{Fairhurst}}},\ }\href {\doibase 10.1103/PhysRevD.103.024042} {\bibfield  {journal} {\bibinfo  {journal} {\prd}\ }\textbf {\bibinfo {volume} {103}},\ \bibinfo {eid} {024042} (\bibinfo {year} {2021})},\ \Eprint {http://arxiv.org/abs/2007.04313} {arXiv:2007.04313 [gr-qc]} \BibitemShut {NoStop}%
\bibitem [{\citenamefont {Harry}\ \emph {et~al.}(2018)\citenamefont {Harry}, \citenamefont {Bustillo},\ and\ \citenamefont {Nitz}}]{Har18}%
  \BibitemOpen
  \bibfield  {author} {\bibinfo {author} {\bibfnamefont {I.}~\bibnamefont {Harry}}, \bibinfo {author} {\bibfnamefont {J.~C.}\ \bibnamefont {Bustillo}}, \ and\ \bibinfo {author} {\bibfnamefont {A.}~\bibnamefont {Nitz}},\ }\href {\doibase 10.1103/PhysRevD.97.023004} {\bibfield  {journal} {\bibinfo  {journal} {Phys. Rev. D}\ }\textbf {\bibinfo {volume} {97}},\ \bibinfo {pages} {023004} (\bibinfo {year} {2018})}\BibitemShut {NoStop}%
\bibitem [{\citenamefont {Chandra}\ \emph {et~al.}(2022)\citenamefont {Chandra}, \citenamefont {Bustillo}, \citenamefont {Pai},\ and\ \citenamefont {Harry}}]{Cha22}%
  \BibitemOpen
  \bibfield  {author} {\bibinfo {author} {\bibfnamefont {K.}~\bibnamefont {Chandra}}, \bibinfo {author} {\bibfnamefont {J.~C.}\ \bibnamefont {Bustillo}}, \bibinfo {author} {\bibfnamefont {A.}~\bibnamefont {Pai}}, \ and\ \bibinfo {author} {\bibfnamefont {I.~W.}\ \bibnamefont {Harry}},\ }\href {\doibase 10.1103/PhysRevD.106.123003} {\bibfield  {journal} {\bibinfo  {journal} {Phys. Rev. D}\ }\textbf {\bibinfo {volume} {106}},\ \bibinfo {pages} {123003} (\bibinfo {year} {2022})}\BibitemShut {NoStop}%
\bibitem [{\citenamefont {{Sharma}}\ \emph {et~al.}(2022)\citenamefont {{Sharma}}, \citenamefont {{Chandra}},\ and\ \citenamefont {{Pai}}}]{Sha22}%
  \BibitemOpen
  \bibfield  {author} {\bibinfo {author} {\bibfnamefont {K.}~\bibnamefont {{Sharma}}}, \bibinfo {author} {\bibfnamefont {K.}~\bibnamefont {{Chandra}}}, \ and\ \bibinfo {author} {\bibfnamefont {A.}~\bibnamefont {{Pai}}},\ }\href@noop {} {\bibfield  {journal} {\bibinfo  {journal} {arXiv e-prints}\ ,\ \bibinfo {eid} {arXiv:2208.02545}} (\bibinfo {year} {2022})},\ \Eprint {http://arxiv.org/abs/2208.02545} {arXiv:2208.02545 [astro-ph.HE]} \BibitemShut {NoStop}%
\bibitem [{\citenamefont {{Zhang}}\ \emph {et~al.}(2023)\citenamefont {{Zhang}}, \citenamefont {{Dai}},\ and\ \citenamefont {{Liang}}}]{Zha23}%
  \BibitemOpen
  \bibfield  {author} {\bibinfo {author} {\bibfnamefont {C.}~\bibnamefont {{Zhang}}}, \bibinfo {author} {\bibfnamefont {N.}~\bibnamefont {{Dai}}}, \ and\ \bibinfo {author} {\bibfnamefont {D.}~\bibnamefont {{Liang}}},\ }\href {\doibase 10.1103/PhysRevD.108.044076} {\bibfield  {journal} {\bibinfo  {journal} {\prd}\ }\textbf {\bibinfo {volume} {108}},\ \bibinfo {eid} {044076} (\bibinfo {year} {2023})},\ \Eprint {http://arxiv.org/abs/2306.13871} {arXiv:2306.13871 [gr-qc]} \BibitemShut {NoStop}%
\bibitem [{\citenamefont {Abbott}\ \emph {et~al.}(2022)\citenamefont {Abbott} \emph {et~al.}}]{LVK_O3_IMBH_search}%
  \BibitemOpen
  \bibfield  {author} {\bibinfo {author} {\bibfnamefont {R.}~\bibnamefont {Abbott}} \emph {et~al.} (\bibinfo {collaboration} {LIGO Scientific, VIRGO, KAGRA}),\ }\href {\doibase 10.1051/0004-6361/202141452} {\bibfield  {journal} {\bibinfo  {journal} {Astron. Astrophys.}\ }\textbf {\bibinfo {volume} {659}},\ \bibinfo {pages} {A84} (\bibinfo {year} {2022})},\ \Eprint {http://arxiv.org/abs/2105.15120} {arXiv:2105.15120 [astro-ph.HE]} \BibitemShut {NoStop}%
\bibitem [{\citenamefont {Coleman~Miller}\ and\ \citenamefont {Colbert}(2004)}]{doi:10.1142/S0218271804004426}%
  \BibitemOpen
  \bibfield  {author} {\bibinfo {author} {\bibfnamefont {M.}~\bibnamefont {Coleman~Miller}}\ and\ \bibinfo {author} {\bibfnamefont {E.~J.~M.}\ \bibnamefont {Colbert}},\ }\href {\doibase 10.1142/S0218271804004426} {\bibfield  {journal} {\bibinfo  {journal} {International Journal of Modern Physics D}\ }\textbf {\bibinfo {volume} {13}},\ \bibinfo {pages} {1} (\bibinfo {year} {2004})},\ \Eprint {http://arxiv.org/abs/https://doi.org/10.1142/S0218271804004426} {https://doi.org/10.1142/S0218271804004426} \BibitemShut {NoStop}%
\bibitem [{\citenamefont {{Gair}}\ \emph {et~al.}(2011)\citenamefont {{Gair}}, \citenamefont {{Mandel}}, \citenamefont {{Miller}},\ and\ \citenamefont {{Volonteri}}}]{2011GReGr..43..485G}%
  \BibitemOpen
  \bibfield  {author} {\bibinfo {author} {\bibfnamefont {J.~R.}\ \bibnamefont {{Gair}}}, \bibinfo {author} {\bibfnamefont {I.}~\bibnamefont {{Mandel}}}, \bibinfo {author} {\bibfnamefont {M.~C.}\ \bibnamefont {{Miller}}}, \ and\ \bibinfo {author} {\bibfnamefont {M.}~\bibnamefont {{Volonteri}}},\ }\href {\doibase 10.1007/s10714-010-1104-3} {\bibfield  {journal} {\bibinfo  {journal} {General Relativity and Gravitation}\ }\textbf {\bibinfo {volume} {43}},\ \bibinfo {pages} {485} (\bibinfo {year} {2011})},\ \Eprint {http://arxiv.org/abs/0907.5450} {arXiv:0907.5450 [astro-ph.CO]} \BibitemShut {NoStop}%
\bibitem [{\citenamefont {{Abbott}}\ \emph {et~al.}(2020{\natexlab{a}})\citenamefont {{Abbott}}, \citenamefont {{LIGO Scientific Collaboration}},\ and\ \citenamefont {{Virgo Collaboration}}}]{GW190412}%
  \BibitemOpen
  \bibfield  {author} {\bibinfo {author} {\bibfnamefont {R.}~\bibnamefont {{Abbott}}}, \bibinfo {author} {\bibnamefont {{LIGO Scientific Collaboration}}}, \ and\ \bibinfo {author} {\bibnamefont {{Virgo Collaboration}}},\ }\href {\doibase 10.1103/PhysRevD.102.043015} {\bibfield  {journal} {\bibinfo  {journal} {\prd}\ }\textbf {\bibinfo {volume} {102}},\ \bibinfo {eid} {043015} (\bibinfo {year} {2020}{\natexlab{a}})},\ \Eprint {http://arxiv.org/abs/2004.08342} {arXiv:2004.08342 [astro-ph.HE]} \BibitemShut {NoStop}%
\bibitem [{\citenamefont {{Abbott}}\ \emph {et~al.}(2020{\natexlab{b}})\citenamefont {{Abbott}} \emph {et~al.}}]{GW190814}%
  \BibitemOpen
  \bibfield  {author} {\bibinfo {author} {\bibfnamefont {R.}~\bibnamefont {{Abbott}}} \emph {et~al.},\ }\href {\doibase 10.3847/2041-8213/ab960f} {\bibfield  {journal} {\bibinfo  {journal} {\apjl}\ }\textbf {\bibinfo {volume} {896}},\ \bibinfo {eid} {L44} (\bibinfo {year} {2020}{\natexlab{b}})},\ \Eprint {http://arxiv.org/abs/2006.12611} {arXiv:2006.12611 [astro-ph.HE]} \BibitemShut {NoStop}%
\bibitem [{\citenamefont {Anagnostou}\ \emph {et~al.}(2022)\citenamefont {Anagnostou}, \citenamefont {Trenti},\ and\ \citenamefont {Melatos}}]{hierarchical_7merger_scenario2020b}%
  \BibitemOpen
  \bibfield  {author} {\bibinfo {author} {\bibfnamefont {O.}~\bibnamefont {Anagnostou}}, \bibinfo {author} {\bibfnamefont {M.}~\bibnamefont {Trenti}}, \ and\ \bibinfo {author} {\bibfnamefont {A.}~\bibnamefont {Melatos}},\ }\href {\doibase 10.3847/1538-4357/ac9d95} {\bibfield  {journal} {\bibinfo  {journal} {The Astrophysical Journal}\ }\textbf {\bibinfo {volume} {941}},\ \bibinfo {pages} {4} (\bibinfo {year} {2022})}\BibitemShut {NoStop}%
\bibitem [{\citenamefont {Fragione}\ \emph {et~al.}(2020)\citenamefont {Fragione}, \citenamefont {Loeb},\ and\ \citenamefont {Rasio}}]{hierarchical_from_dynamical_in_any_star_cluster2020b}%
  \BibitemOpen
  \bibfield  {author} {\bibinfo {author} {\bibfnamefont {G.}~\bibnamefont {Fragione}}, \bibinfo {author} {\bibfnamefont {A.}~\bibnamefont {Loeb}}, \ and\ \bibinfo {author} {\bibfnamefont {F.~A.}\ \bibnamefont {Rasio}},\ }\href {\doibase 10.3847/2041-8213/abbc0a} {\bibfield  {journal} {\bibinfo  {journal} {Astrophys. J. Lett.}\ }\textbf {\bibinfo {volume} {902}},\ \bibinfo {pages} {L26} (\bibinfo {year} {2020})},\ \Eprint {http://arxiv.org/abs/2009.05065} {arXiv:2009.05065 [astro-ph.GA]} \BibitemShut {NoStop}%
\bibitem [{\citenamefont {Veske}\ \emph {et~al.}(2021)\citenamefont {Veske}, \citenamefont {Sullivan}, \citenamefont {M\'arka}, \citenamefont {Bartos}, \citenamefont {Corley}, \citenamefont {Samsing}, \citenamefont {Buscicchio},\ and\ \citenamefont {M\'arka}}]{hierarchical_mergerFamily_dynamical_mass_dist_matters2021}%
  \BibitemOpen
  \bibfield  {author} {\bibinfo {author} {\bibfnamefont {D.}~\bibnamefont {Veske}}, \bibinfo {author} {\bibfnamefont {A.~G.}\ \bibnamefont {Sullivan}}, \bibinfo {author} {\bibfnamefont {Z.}~\bibnamefont {M\'arka}}, \bibinfo {author} {\bibfnamefont {I.}~\bibnamefont {Bartos}},  \emph {et~al.},\ }\href {\doibase 10.3847/2041-8213/abd721} {\bibfield  {journal} {\bibinfo  {journal} {Astrophys. J. Lett.}\ }\textbf {\bibinfo {volume} {907}},\ \bibinfo {pages} {L48} (\bibinfo {year} {2021})},\ \Eprint {http://arxiv.org/abs/2011.06591} {arXiv:2011.06591 [astro-ph.HE]} \BibitemShut {NoStop}%
\bibitem [{\citenamefont {Fragione}\ \emph {et~al.}(2022)\citenamefont {Fragione}, \citenamefont {Kocsis}, \citenamefont {Rasio},\ and\ \citenamefont {Silk}}]{hierarchical_rate_sensitive_to_natal_spins_Fragione2021kocsis}%
  \BibitemOpen
  \bibfield  {author} {\bibinfo {author} {\bibfnamefont {G.}~\bibnamefont {Fragione}}, \bibinfo {author} {\bibfnamefont {B.}~\bibnamefont {Kocsis}}, \bibinfo {author} {\bibfnamefont {F.~A.}\ \bibnamefont {Rasio}}, \ and\ \bibinfo {author} {\bibfnamefont {J.}~\bibnamefont {Silk}},\ }\href {\doibase 10.3847/1538-4357/ac5026} {\bibfield  {journal} {\bibinfo  {journal} {The Astrophysical Journal}\ }\textbf {\bibinfo {volume} {927}},\ \bibinfo {pages} {231} (\bibinfo {year} {2022})}\BibitemShut {NoStop}%
\bibitem [{\citenamefont {Yang}\ \emph {et~al.}(2019)\citenamefont {Yang}, \citenamefont {Bartos}, \citenamefont {Gayathri}, \citenamefont {Ford}, \citenamefont {Haiman}, \citenamefont {Klimenko}, \citenamefont {Kocsis}, \citenamefont {M\'arka}, \citenamefont {M\'arka}, \citenamefont {McKernan},\ and\ \citenamefont {O'Shaughnessy}}]{hierarchical_mergers_agn_kocsis2019}%
  \BibitemOpen
  \bibfield  {author} {\bibinfo {author} {\bibfnamefont {Y.}~\bibnamefont {Yang}}, \bibinfo {author} {\bibfnamefont {I.}~\bibnamefont {Bartos}}, \bibinfo {author} {\bibfnamefont {V.}~\bibnamefont {Gayathri}}, \bibinfo {author} {\bibfnamefont {K.~E.~S.}\ \bibnamefont {Ford}},  \emph {et~al.},\ }\href {\doibase 10.1103/PhysRevLett.123.181101} {\bibfield  {journal} {\bibinfo  {journal} {Phys. Rev. Lett.}\ }\textbf {\bibinfo {volume} {123}},\ \bibinfo {pages} {181101} (\bibinfo {year} {2019})}\BibitemShut {NoStop}%
\bibitem [{\citenamefont {McKernan}\ \emph {et~al.}(2020)\citenamefont {McKernan}, \citenamefont {Ford}, \citenamefont {O’Shaugnessy},\ and\ \citenamefont {Wysocki}}]{agn_bbh_population_chieff_q_simulation_mckernan_ford2019}%
  \BibitemOpen
  \bibfield  {author} {\bibinfo {author} {\bibfnamefont {B.}~\bibnamefont {McKernan}}, \bibinfo {author} {\bibfnamefont {K.~E.~S.}\ \bibnamefont {Ford}}, \bibinfo {author} {\bibfnamefont {R.}~\bibnamefont {O’Shaugnessy}}, \ and\ \bibinfo {author} {\bibfnamefont {D.}~\bibnamefont {Wysocki}},\ }\href {\doibase 10.1093/mnras/staa740} {\bibfield  {journal} {\bibinfo  {journal} {Monthly Notices of the Royal Astronomical Society}\ }\textbf {\bibinfo {volume} {494}},\ \bibinfo {pages} {1203} (\bibinfo {year} {2020})},\ \Eprint {http://arxiv.org/abs/https://academic.oup.com/mnras/article-pdf/494/1/1203/33029648/staa740.pdf} {https://academic.oup.com/mnras/article-pdf/494/1/1203/33029648/staa740.pdf} \BibitemShut {NoStop}%
\bibitem [{\citenamefont {Tagawa}\ \emph {et~al.}(2020)\citenamefont {Tagawa}, \citenamefont {Haiman}, \citenamefont {Bartos},\ and\ \citenamefont {Kocsis}}]{bbh_spin_evolution_agn_Tagawa_2020a}%
  \BibitemOpen
  \bibfield  {author} {\bibinfo {author} {\bibfnamefont {H.}~\bibnamefont {Tagawa}}, \bibinfo {author} {\bibfnamefont {Z.}~\bibnamefont {Haiman}}, \bibinfo {author} {\bibfnamefont {I.}~\bibnamefont {Bartos}}, \ and\ \bibinfo {author} {\bibfnamefont {B.}~\bibnamefont {Kocsis}},\ }\href {\doibase 10.3847/1538-4357/aba2cc} {\bibfield  {journal} {\bibinfo  {journal} {The Astrophysical Journal}\ }\textbf {\bibinfo {volume} {899}},\ \bibinfo {pages} {26} (\bibinfo {year} {2020})}\BibitemShut {NoStop}%
\bibitem [{\citenamefont {{Ishibashi, W.}}\ and\ \citenamefont {{Gr\"obner, M.}}(2020)}]{bbh_evolution_agn_merger_timescale_ishibashi2020a}%
  \BibitemOpen
  \bibfield  {author} {\bibinfo {author} {\bibnamefont {{Ishibashi, W.}}}\ and\ \bibinfo {author} {\bibnamefont {{Gr\"obner, M.}}},\ }\href {\doibase 10.1051/0004-6361/202037799} {\bibfield  {journal} {\bibinfo  {journal} {A\&A}\ }\textbf {\bibinfo {volume} {639}},\ \bibinfo {pages} {A108} (\bibinfo {year} {2020})}\BibitemShut {NoStop}%
\bibitem [{\citenamefont {Secunda}\ \emph {et~al.}(2020)\citenamefont {Secunda}, \citenamefont {Bellovary}, \citenamefont {Mac~Low}, \citenamefont {Ford}, \citenamefont {McKernan}, \citenamefont {Leigh}, \citenamefont {Lyra}, \citenamefont {Sandor},\ and\ \citenamefont {Adorno}}]{migration_traps_spins_rates_mckernan_ford2020a}%
  \BibitemOpen
  \bibfield  {author} {\bibinfo {author} {\bibfnamefont {A.}~\bibnamefont {Secunda}}, \bibinfo {author} {\bibfnamefont {J.}~\bibnamefont {Bellovary}}, \bibinfo {author} {\bibfnamefont {M.-M.}\ \bibnamefont {Mac~Low}}, \bibinfo {author} {\bibfnamefont {K.~E.~S.}\ \bibnamefont {Ford}},  \emph {et~al.},\ }\href {\doibase 10.3847/1538-4357/abbc1d} {\bibfield  {journal} {\bibinfo  {journal} {Astrophys. J.}\ }\textbf {\bibinfo {volume} {903}},\ \bibinfo {pages} {133} (\bibinfo {year} {2020})},\ \Eprint {http://arxiv.org/abs/2004.11936} {arXiv:2004.11936 [astro-ph.HE]} \BibitemShut {NoStop}%
\bibitem [{\citenamefont {Tagawa}\ \emph {et~al.}(2021)\citenamefont {Tagawa}, \citenamefont {Kocsis}, \citenamefont {Haiman}, \citenamefont {Bartos}, \citenamefont {Omukai},\ and\ \citenamefont {Samsing}}]{mass_gap_agn_bbh_mergers2021}%
  \BibitemOpen
  \bibfield  {author} {\bibinfo {author} {\bibfnamefont {H.}~\bibnamefont {Tagawa}}, \bibinfo {author} {\bibfnamefont {B.}~\bibnamefont {Kocsis}}, \bibinfo {author} {\bibfnamefont {Z.}~\bibnamefont {Haiman}}, \bibinfo {author} {\bibfnamefont {I.}~\bibnamefont {Bartos}}, \bibinfo {author} {\bibfnamefont {K.}~\bibnamefont {Omukai}}, \ and\ \bibinfo {author} {\bibfnamefont {J.}~\bibnamefont {Samsing}},\ }\href {\doibase 10.3847/1538-4357/abd555} {\bibfield  {journal} {\bibinfo  {journal} {Astrophys. J.}\ }\textbf {\bibinfo {volume} {908}},\ \bibinfo {pages} {194} (\bibinfo {year} {2021})},\ \Eprint {http://arxiv.org/abs/2012.00011} {arXiv:2012.00011 [astro-ph.HE]} \BibitemShut {NoStop}%
\bibitem [{\citenamefont {Fabj}\ \emph {et~al.}(2020)\citenamefont {Fabj}, \citenamefont {Nasim}, \citenamefont {Caban}, \citenamefont {Ford}, \citenamefont {McKernan},\ and\ \citenamefont {Bellovary}}]{agn_accretion_disk_merger_population2020a}%
  \BibitemOpen
  \bibfield  {author} {\bibinfo {author} {\bibfnamefont {G.}~\bibnamefont {Fabj}}, \bibinfo {author} {\bibfnamefont {S.~S.}\ \bibnamefont {Nasim}}, \bibinfo {author} {\bibfnamefont {F.}~\bibnamefont {Caban}}, \bibinfo {author} {\bibfnamefont {K.~E.~S.}\ \bibnamefont {Ford}}, \bibinfo {author} {\bibfnamefont {B.}~\bibnamefont {McKernan}}, \ and\ \bibinfo {author} {\bibfnamefont {J.~M.}\ \bibnamefont {Bellovary}},\ }\href {\doibase 10.1093/mnras/staa3004} {\bibfield  {journal} {\bibinfo  {journal} {Mon. Not. Roy. Astron. Soc.}\ }\textbf {\bibinfo {volume} {499}},\ \bibinfo {pages} {2608} (\bibinfo {year} {2020})},\ \Eprint {http://arxiv.org/abs/2006.11229} {arXiv:2006.11229 [astro-ph.GA]} \BibitemShut {NoStop}%
\bibitem [{\citenamefont {{Tagawa}}\ \emph {et~al.}(2020)\citenamefont {{Tagawa}}, \citenamefont {{Haiman}}, \citenamefont {{Bartos}},\ and\ \citenamefont {{Kocsis}}}]{Tag20_AGN}%
  \BibitemOpen
  \bibfield  {author} {\bibinfo {author} {\bibfnamefont {H.}~\bibnamefont {{Tagawa}}}, \bibinfo {author} {\bibfnamefont {Z.}~\bibnamefont {{Haiman}}}, \bibinfo {author} {\bibfnamefont {I.}~\bibnamefont {{Bartos}}}, \ and\ \bibinfo {author} {\bibfnamefont {B.}~\bibnamefont {{Kocsis}}},\ }\href {\doibase 10.3847/1538-4357/aba2cc} {\bibfield  {journal} {\bibinfo  {journal} {\apj}\ }\textbf {\bibinfo {volume} {899}},\ \bibinfo {eid} {26} (\bibinfo {year} {2020})},\ \Eprint {http://arxiv.org/abs/2004.11914} {arXiv:2004.11914 [astro-ph.HE]} \BibitemShut {NoStop}%
\bibitem [{\citenamefont {{Samsing}}\ \emph {et~al.}(2022)\citenamefont {{Samsing}}, \citenamefont {{Bartos}}, \citenamefont {{D'Orazio}}, \citenamefont {{Haiman}}, \citenamefont {{Kocsis}}, \citenamefont {{Leigh}}, \citenamefont {{Liu}}, \citenamefont {{Pessah}},\ and\ \citenamefont {{Tagawa}}}]{Sam22_AGN}%
  \BibitemOpen
  \bibfield  {author} {\bibinfo {author} {\bibfnamefont {J.}~\bibnamefont {{Samsing}}}, \bibinfo {author} {\bibfnamefont {I.}~\bibnamefont {{Bartos}}}, \bibinfo {author} {\bibfnamefont {D.~J.}\ \bibnamefont {{D'Orazio}}}, \bibinfo {author} {\bibfnamefont {Z.}~\bibnamefont {{Haiman}}},  \emph {et~al.},\ }\href {\doibase 10.1038/s41586-021-04333-1} {\bibfield  {journal} {\bibinfo  {journal} {\nat}\ }\textbf {\bibinfo {volume} {603}},\ \bibinfo {pages} {237} (\bibinfo {year} {2022})},\ \Eprint {http://arxiv.org/abs/2010.09765} {arXiv:2010.09765 [astro-ph.HE]} \BibitemShut {NoStop}%
\bibitem [{\citenamefont {Marchant}\ and\ \citenamefont {Moriya}(2020)}]{Marchant:2020haw}%
  \BibitemOpen
  \bibfield  {author} {\bibinfo {author} {\bibfnamefont {P.}~\bibnamefont {Marchant}}\ and\ \bibinfo {author} {\bibfnamefont {T.}~\bibnamefont {Moriya}},\ }\href {\doibase 10.1051/0004-6361/202038902} {\bibfield  {journal} {\bibinfo  {journal} {Astron. Astrophys.}\ }\textbf {\bibinfo {volume} {640}},\ \bibinfo {pages} {L18} (\bibinfo {year} {2020})},\ \Eprint {http://arxiv.org/abs/2007.06220} {arXiv:2007.06220 [astro-ph.HE]} \BibitemShut {NoStop}%
\bibitem [{\citenamefont {Farmer}\ \emph {et~al.}(2019)\citenamefont {Farmer}, \citenamefont {Renzo}, \citenamefont {de~Mink}, \citenamefont {Marchant},\ and\ \citenamefont {Justham}}]{Farmer:2019jed}%
  \BibitemOpen
  \bibfield  {author} {\bibinfo {author} {\bibfnamefont {R.}~\bibnamefont {Farmer}}, \bibinfo {author} {\bibfnamefont {M.}~\bibnamefont {Renzo}}, \bibinfo {author} {\bibfnamefont {S.~E.}\ \bibnamefont {de~Mink}}, \bibinfo {author} {\bibfnamefont {P.}~\bibnamefont {Marchant}}, \ and\ \bibinfo {author} {\bibfnamefont {S.}~\bibnamefont {Justham}},\ }\href {\doibase 10.3847/1538-4357/ab518b} {\  (\bibinfo {year} {2019}),\ 10.3847/1538-4357/ab518b},\ \Eprint {http://arxiv.org/abs/1910.12874} {arXiv:1910.12874 [astro-ph.SR]} \BibitemShut {NoStop}%
\bibitem [{\citenamefont {{van Son}}\ \emph {et~al.}(2020)\citenamefont {{van Son}}, \citenamefont {{De Mink}}, \citenamefont {{Broekgaarden}}, \citenamefont {{Renzo}}, \citenamefont {{Justham}}, \citenamefont {{Laplace}}, \citenamefont {{Mor{\'a}n-Fraile}}, \citenamefont {{Hendriks}},\ and\ \citenamefont {{Farmer}}}]{VanSon20_UMG_Pollution}%
  \BibitemOpen
  \bibfield  {author} {\bibinfo {author} {\bibfnamefont {L.~A.~C.}\ \bibnamefont {{van Son}}}, \bibinfo {author} {\bibfnamefont {S.~E.}\ \bibnamefont {{De Mink}}}, \bibinfo {author} {\bibfnamefont {F.~S.}\ \bibnamefont {{Broekgaarden}}}, \bibinfo {author} {\bibfnamefont {M.}~\bibnamefont {{Renzo}}},  \emph {et~al.},\ }\href {\doibase 10.3847/1538-4357/ab9809} {\bibfield  {journal} {\bibinfo  {journal} {\apj}\ }\textbf {\bibinfo {volume} {897}},\ \bibinfo {eid} {100} (\bibinfo {year} {2020})},\ \Eprint {http://arxiv.org/abs/2004.05187} {arXiv:2004.05187 [astro-ph.HE]} \BibitemShut {NoStop}%
\bibitem [{\citenamefont {Mehta}\ \emph {et~al.}(2022)\citenamefont {Mehta}, \citenamefont {Buonanno}, \citenamefont {Gair}, \citenamefont {Miller}, \citenamefont {Farag}, \citenamefont {deBoer}, \citenamefont {Wiescher},\ and\ \citenamefont {Timmes}}]{Mehta:2021fgz}%
  \BibitemOpen
  \bibfield  {author} {\bibinfo {author} {\bibfnamefont {A.~K.}\ \bibnamefont {Mehta}}, \bibinfo {author} {\bibfnamefont {A.}~\bibnamefont {Buonanno}}, \bibinfo {author} {\bibfnamefont {J.}~\bibnamefont {Gair}}, \bibinfo {author} {\bibfnamefont {M.~C.}\ \bibnamefont {Miller}},  \emph {et~al.},\ }\href {\doibase 10.3847/1538-4357/ac3130} {\bibfield  {journal} {\bibinfo  {journal} {Astrophys. J.}\ }\textbf {\bibinfo {volume} {924}},\ \bibinfo {pages} {39} (\bibinfo {year} {2022})},\ \Eprint {http://arxiv.org/abs/2105.06366} {arXiv:2105.06366 [gr-qc]} \BibitemShut {NoStop}%
\bibitem [{\citenamefont {{Hendriks}}\ \emph {et~al.}(2023)\citenamefont {{Hendriks}}, \citenamefont {{van Son}}, \citenamefont {{Renzo}}, \citenamefont {{Izzard}},\ and\ \citenamefont {{Farmer}}}]{Hen23_UMG}%
  \BibitemOpen
  \bibfield  {author} {\bibinfo {author} {\bibfnamefont {D.~D.}\ \bibnamefont {{Hendriks}}}, \bibinfo {author} {\bibfnamefont {L.~A.~C.}\ \bibnamefont {{van Son}}}, \bibinfo {author} {\bibfnamefont {M.}~\bibnamefont {{Renzo}}}, \bibinfo {author} {\bibfnamefont {R.~G.}\ \bibnamefont {{Izzard}}}, \ and\ \bibinfo {author} {\bibfnamefont {R.}~\bibnamefont {{Farmer}}},\ }\href {\doibase 10.1093/mnras/stad2857} {\bibfield  {journal} {\bibinfo  {journal} {\mnras}\ } (\bibinfo {year} {2023}),\ 10.1093/mnras/stad2857},\ \Eprint {http://arxiv.org/abs/2309.09339} {arXiv:2309.09339 [astro-ph.HE]} \BibitemShut {NoStop}%
\bibitem [{\citenamefont {Golomb}\ \emph {et~al.}(2024)\citenamefont {Golomb}, \citenamefont {Isi},\ and\ \citenamefont {Farr}}]{Gol23_UMG}%
  \BibitemOpen
  \bibfield  {author} {\bibinfo {author} {\bibfnamefont {J.}~\bibnamefont {Golomb}}, \bibinfo {author} {\bibfnamefont {M.}~\bibnamefont {Isi}}, \ and\ \bibinfo {author} {\bibfnamefont {W.~M.}\ \bibnamefont {Farr}},\ }\href {\doibase 10.3847/1538-4357/ad8572} {\bibfield  {journal} {\bibinfo  {journal} {The Astrophysical Journal}\ }\textbf {\bibinfo {volume} {976}},\ \bibinfo {pages} {121} (\bibinfo {year} {2024})}\BibitemShut {NoStop}%
\bibitem [{\citenamefont {{Franciolini}}\ \emph {et~al.}(2024)\citenamefont {{Franciolini}}, \citenamefont {{Kritos}}, \citenamefont {{Reali}}, \citenamefont {{Broekgaarden}},\ and\ \citenamefont {{Berti}}}]{Fra24_UMG}%
  \BibitemOpen
  \bibfield  {author} {\bibinfo {author} {\bibfnamefont {G.}~\bibnamefont {{Franciolini}}}, \bibinfo {author} {\bibfnamefont {K.}~\bibnamefont {{Kritos}}}, \bibinfo {author} {\bibfnamefont {L.}~\bibnamefont {{Reali}}}, \bibinfo {author} {\bibfnamefont {F.}~\bibnamefont {{Broekgaarden}}}, \ and\ \bibinfo {author} {\bibfnamefont {E.}~\bibnamefont {{Berti}}},\ }\href {\doibase 10.48550/arXiv.2401.13038} {\bibfield  {journal} {\bibinfo  {journal} {arXiv e-prints}\ ,\ \bibinfo {eid} {arXiv:2401.13038}} (\bibinfo {year} {2024})},\ \Eprint {http://arxiv.org/abs/2401.13038} {arXiv:2401.13038 [astro-ph.HE]} \BibitemShut {NoStop}%
\bibitem [{\citenamefont {{Fragione}}\ \emph {et~al.}(2018)\citenamefont {{Fragione}}, \citenamefont {{Ginsburg}},\ and\ \citenamefont {{Kocsis}}}]{Fra18_IMBH_LISA}%
  \BibitemOpen
  \bibfield  {author} {\bibinfo {author} {\bibfnamefont {G.}~\bibnamefont {{Fragione}}}, \bibinfo {author} {\bibfnamefont {I.}~\bibnamefont {{Ginsburg}}}, \ and\ \bibinfo {author} {\bibfnamefont {B.}~\bibnamefont {{Kocsis}}},\ }\href {\doibase 10.3847/1538-4357/aab368} {\bibfield  {journal} {\bibinfo  {journal} {\apj}\ }\textbf {\bibinfo {volume} {856}},\ \bibinfo {eid} {92} (\bibinfo {year} {2018})},\ \Eprint {http://arxiv.org/abs/1711.00483} {arXiv:1711.00483 [astro-ph.GA]} \BibitemShut {NoStop}%
\bibitem [{\citenamefont {{Kritos}}\ \emph {et~al.}(2023)\citenamefont {{Kritos}}, \citenamefont {{Berti}},\ and\ \citenamefont {{Silk}}}]{Kri23_NSC}%
  \BibitemOpen
  \bibfield  {author} {\bibinfo {author} {\bibfnamefont {K.}~\bibnamefont {{Kritos}}}, \bibinfo {author} {\bibfnamefont {E.}~\bibnamefont {{Berti}}}, \ and\ \bibinfo {author} {\bibfnamefont {J.}~\bibnamefont {{Silk}}},\ }\href {\doibase 10.1103/PhysRevD.108.083012} {\bibfield  {journal} {\bibinfo  {journal} {\prd}\ }\textbf {\bibinfo {volume} {108}},\ \bibinfo {eid} {083012} (\bibinfo {year} {2023})},\ \Eprint {http://arxiv.org/abs/2212.06845} {arXiv:2212.06845 [astro-ph.HE]} \BibitemShut {NoStop}%
\bibitem [{\citenamefont {{Kritos}}\ \emph {et~al.}(2024{\natexlab{a}})\citenamefont {{Kritos}}, \citenamefont {{Reali}}, \citenamefont {{Gerosa}},\ and\ \citenamefont {{Berti}}}]{Kri24_GasAccretion}%
  \BibitemOpen
  \bibfield  {author} {\bibinfo {author} {\bibfnamefont {K.}~\bibnamefont {{Kritos}}}, \bibinfo {author} {\bibfnamefont {L.}~\bibnamefont {{Reali}}}, \bibinfo {author} {\bibfnamefont {D.}~\bibnamefont {{Gerosa}}}, \ and\ \bibinfo {author} {\bibfnamefont {E.}~\bibnamefont {{Berti}}},\ }\href {\doibase 10.1103/PhysRevD.110.123017} {\bibfield  {journal} {\bibinfo  {journal} {\prd}\ }\textbf {\bibinfo {volume} {110}},\ \bibinfo {eid} {123017} (\bibinfo {year} {2024}{\natexlab{a}})},\ \Eprint {http://arxiv.org/abs/2409.15439} {arXiv:2409.15439 [astro-ph.HE]} \BibitemShut {NoStop}%
\bibitem [{\citenamefont {{Kritos}}\ \emph {et~al.}(2024{\natexlab{b}})\citenamefont {{Kritos}}, \citenamefont {{Strokov}}, \citenamefont {{Baibhav}},\ and\ \citenamefont {{Berti}}}]{Kri24_Rapster}%
  \BibitemOpen
  \bibfield  {author} {\bibinfo {author} {\bibfnamefont {K.}~\bibnamefont {{Kritos}}}, \bibinfo {author} {\bibfnamefont {V.}~\bibnamefont {{Strokov}}}, \bibinfo {author} {\bibfnamefont {V.}~\bibnamefont {{Baibhav}}}, \ and\ \bibinfo {author} {\bibfnamefont {E.}~\bibnamefont {{Berti}}},\ }\href {\doibase 10.1103/PhysRevD.110.043023} {\bibfield  {journal} {\bibinfo  {journal} {\prd}\ }\textbf {\bibinfo {volume} {110}},\ \bibinfo {eid} {043023} (\bibinfo {year} {2024}{\natexlab{b}})},\ \Eprint {http://arxiv.org/abs/2210.10055} {arXiv:2210.10055 [astro-ph.HE]} \BibitemShut {NoStop}%
\bibitem [{\citenamefont {Chandra}\ \emph {et~al.}(2020)\citenamefont {Chandra}, \citenamefont {Gayathri}, \citenamefont {Bustillo},\ and\ \citenamefont {Pai}}]{Cha20_VT}%
  \BibitemOpen
  \bibfield  {author} {\bibinfo {author} {\bibfnamefont {K.}~\bibnamefont {Chandra}}, \bibinfo {author} {\bibfnamefont {V.}~\bibnamefont {Gayathri}}, \bibinfo {author} {\bibfnamefont {J.~C.}\ \bibnamefont {Bustillo}}, \ and\ \bibinfo {author} {\bibfnamefont {A.}~\bibnamefont {Pai}},\ }\href {\doibase 10.1103/PhysRevD.102.044035} {\bibfield  {journal} {\bibinfo  {journal} {Phys. Rev. D}\ }\textbf {\bibinfo {volume} {102}},\ \bibinfo {pages} {044035} (\bibinfo {year} {2020})}\BibitemShut {NoStop}%
\bibitem [{\citenamefont {{Fairhurst}}\ \emph {et~al.}(2023)\citenamefont {{Fairhurst}}, \citenamefont {{Hoy}}, \citenamefont {{Green}}, \citenamefont {{Mills}},\ and\ \citenamefont {{Usman}}}]{Fai23_SimplePE}%
  \BibitemOpen
  \bibfield  {author} {\bibinfo {author} {\bibfnamefont {S.}~\bibnamefont {{Fairhurst}}}, \bibinfo {author} {\bibfnamefont {C.}~\bibnamefont {{Hoy}}}, \bibinfo {author} {\bibfnamefont {R.}~\bibnamefont {{Green}}}, \bibinfo {author} {\bibfnamefont {C.}~\bibnamefont {{Mills}}}, \ and\ \bibinfo {author} {\bibfnamefont {S.~A.}\ \bibnamefont {{Usman}}},\ }\href {\doibase 10.1103/PhysRevD.108.082006} {\bibfield  {journal} {\bibinfo  {journal} {\prd}\ }\textbf {\bibinfo {volume} {108}},\ \bibinfo {eid} {082006} (\bibinfo {year} {2023})},\ \Eprint {http://arxiv.org/abs/2304.03731} {arXiv:2304.03731 [gr-qc]} \BibitemShut {NoStop}%
\bibitem [{\citenamefont {Roulet}\ \emph {et~al.}(2024)\citenamefont {Roulet}, \citenamefont {Mushkin}, \citenamefont {Wadekar}, \citenamefont {Venumadhav}, \citenamefont {Zackay},\ and\ \citenamefont {Zaldarriaga}}]{Rou23_CoherentScore}%
  \BibitemOpen
  \bibfield  {author} {\bibinfo {author} {\bibfnamefont {J.}~\bibnamefont {Roulet}}, \bibinfo {author} {\bibfnamefont {J.}~\bibnamefont {Mushkin}}, \bibinfo {author} {\bibfnamefont {D.}~\bibnamefont {Wadekar}}, \bibinfo {author} {\bibfnamefont {T.}~\bibnamefont {Venumadhav}}, \bibinfo {author} {\bibfnamefont {B.}~\bibnamefont {Zackay}}, \ and\ \bibinfo {author} {\bibfnamefont {M.}~\bibnamefont {Zaldarriaga}},\ }\href {\doibase 10.1103/PhysRevD.110.044010} {\bibfield  {journal} {\bibinfo  {journal} {Phys. Rev. D}\ }\textbf {\bibinfo {volume} {110}},\ \bibinfo {pages} {044010} (\bibinfo {year} {2024})}\BibitemShut {NoStop}%
\bibitem [{\citenamefont {LIGO Scientific~Collaboration}(2021)}]{zenodoLVK}%
  \BibitemOpen
  \bibfield  {author} {\bibinfo {author} {\bibfnamefont {K.~C.}\ \bibnamefont {LIGO Scientific~Collaboration}, \bibfnamefont {Virgo~Collaboration}},\ }\href {\doibase 10.5281/zenodo.5546675} {\  (\bibinfo {year} {2021}),\ 10.5281/zenodo.5546675}\BibitemShut {NoStop}%
\bibitem [{\citenamefont {Tiwari}(2018)}]{Tiwari:2017ndi}%
  \BibitemOpen
  \bibfield  {author} {\bibinfo {author} {\bibfnamefont {V.}~\bibnamefont {Tiwari}},\ }\href {\doibase 10.1088/1361-6382/aac89d} {\bibfield  {journal} {\bibinfo  {journal} {Class. Quant. Grav.}\ }\textbf {\bibinfo {volume} {35}},\ \bibinfo {pages} {145009} (\bibinfo {year} {2018})},\ \Eprint {http://arxiv.org/abs/1712.00482} {arXiv:1712.00482 [astro-ph.HE]} \BibitemShut {NoStop}%
\bibitem [{\citenamefont {Farr}(2019)}]{Farr_2019}%
  \BibitemOpen
  \bibfield  {author} {\bibinfo {author} {\bibfnamefont {W.~M.}\ \bibnamefont {Farr}},\ }\href {\doibase 10.3847/2515-5172/ab1d5f} {\bibfield  {journal} {\bibinfo  {journal} {Research Notes of the AAS}\ }\textbf {\bibinfo {volume} {3}},\ \bibinfo {pages} {66} (\bibinfo {year} {2019})}\BibitemShut {NoStop}%
\bibitem [{\citenamefont {Pratten}\ \emph {et~al.}(2021)\citenamefont {Pratten} \emph {et~al.}}]{Pratten:2020ceb}%
  \BibitemOpen
  \bibfield  {author} {\bibinfo {author} {\bibfnamefont {G.}~\bibnamefont {Pratten}} \emph {et~al.},\ }\href {\doibase 10.1103/PhysRevD.103.104056} {\bibfield  {journal} {\bibinfo  {journal} {Phys. Rev. D}\ }\textbf {\bibinfo {volume} {103}},\ \bibinfo {pages} {104056} (\bibinfo {year} {2021})},\ \Eprint {http://arxiv.org/abs/2004.06503} {arXiv:2004.06503 [gr-qc]} \BibitemShut {NoStop}%
\bibitem [{\citenamefont {Dal~Canton}\ and\ \citenamefont {Harry}(2017)}]{DalCanton:2017ala}%
  \BibitemOpen
  \bibfield  {author} {\bibinfo {author} {\bibfnamefont {T.}~\bibnamefont {Dal~Canton}}\ and\ \bibinfo {author} {\bibfnamefont {I.~W.}\ \bibnamefont {Harry}},\ }\href@noop {} {\  (\bibinfo {year} {2017})},\ \Eprint {http://arxiv.org/abs/1705.01845} {arXiv:1705.01845 [gr-qc]} \BibitemShut {NoStop}%
\bibitem [{\citenamefont {Mehta}(2024)}]{zenodoIAS}%
  \BibitemOpen
  \bibfield  {author} {\bibinfo {author} {\bibfnamefont {A.~K.}\ \bibnamefont {Mehta}},\ }\href {\doibase 10.5281/zenodo.14752873} {\  (\bibinfo {year} {2024}),\ 10.5281/zenodo.14752873}\BibitemShut {NoStop}%
\bibitem [{\citenamefont {Banagiri}\ \emph {et~al.}(2023)\citenamefont {Banagiri}, \citenamefont {Berry}, \citenamefont {Cabourn~Davies}, \citenamefont {Tsukada},\ and\ \citenamefont {Doctor}}]{Banagiri:2023ztt}%
  \BibitemOpen
  \bibfield  {author} {\bibinfo {author} {\bibfnamefont {S.}~\bibnamefont {Banagiri}}, \bibinfo {author} {\bibfnamefont {C.~P.~L.}\ \bibnamefont {Berry}}, \bibinfo {author} {\bibfnamefont {G.~S.}\ \bibnamefont {Cabourn~Davies}}, \bibinfo {author} {\bibfnamefont {L.}~\bibnamefont {Tsukada}}, \ and\ \bibinfo {author} {\bibfnamefont {Z.}~\bibnamefont {Doctor}},\ }\href {\doibase 10.1103/PhysRevD.108.083043} {\bibfield  {journal} {\bibinfo  {journal} {Phys. Rev. D}\ }\textbf {\bibinfo {volume} {108}},\ \bibinfo {pages} {083043} (\bibinfo {year} {2023})},\ \Eprint {http://arxiv.org/abs/2305.00071} {arXiv:2305.00071 [astro-ph.IM]} \BibitemShut {NoStop}%
\bibitem [{\citenamefont {Biswas}\ \emph {et~al.}(2012)\citenamefont {Biswas} \emph {et~al.}}]{Bis12_Combining_Pipelines}%
  \BibitemOpen
  \bibfield  {author} {\bibinfo {author} {\bibfnamefont {R.}~\bibnamefont {Biswas}} \emph {et~al.},\ }\href {\doibase 10.1103/PhysRevD.85.122009} {\bibfield  {journal} {\bibinfo  {journal} {Phys. Rev. D}\ }\textbf {\bibinfo {volume} {85}},\ \bibinfo {pages} {122009} (\bibinfo {year} {2012})},\ \Eprint {http://arxiv.org/abs/1201.2964} {arXiv:1201.2964 [gr-qc]} \BibitemShut {NoStop}%
\end{thebibliography}%
%------------------------------------------------------------------------------

\end{document}